\documentclass[final,12pt,3p,authoryear]{elsarticle/elsarticle}
\journal{Journal of the Taiwan Institute of Chemical Engineers}
\emergencystretch=\maxdimen
\graphicspath{{../Figures/} {Figures/}}
\makeatletter
\def\input@path{ {./styles/} {./misc/} {bibliography/} {./bibliography/} {../bibliography/} }
\makeatother
\newcommand*{\BibPath}{bibliography/}%
\newcommand{\InputBibFiles}[1]{\bibliography{\BibPath#1}}
\usepackage{hyperref}
 \hypersetup{
 	plainpages = false, 
	bookmarksopen = true,
	bookmarksnumbered = true,
	breaklinks = true,
	linktocpage,
	colorlinks = true,
	linkcolor = purple,
	urlcolor  = black,
	citecolor = purple,
	}
\usepackage{multicol}
\usepackage{multirow}
\usepackage{imakeidx}
\usepackage{nomencl}
\makenomenclature
\usepackage[xindy]{glossaries}
\makeglossaries
\usepackage{graphics}
\usepackage{graphicx}
\usepackage{epsf,psfrag}
\usepackage{epstopdf}
\usepackage[para]{threeparttable}
\usepackage{subfig}
\usepackage{epsfig}
\usepackage{pdflscape}
\usepackage{rotating}
\usepackage{lscape}
\usepackage{float}
\usepackage{stfloats}
\usepackage{textgreek}
\usepackage{longtable}
\usepackage{lscape}
\usepackage{capt-of}
\usepackage{comment}
\usepackage{tablefootnote}
\usepackage{footnote}
\usepackage{caption}
\usepackage[autopunct=true]{csquotes}
\usepackage{rotating}
\usepackage{txfonts}
\usepackage[normalem]{ulem}
\usepackage{setspace}
\usepackage{color,xcolor,soul}
\usepackage{amsfonts,amsmath,amssymb}
\usepackage{latexsym,array}
\usepackage{textcomp}

\newcommand{\RomanNumeralCaps}[1]
\linenumbers
\makesavenoteenv{tabular}
\makesavenoteenv{table}
\emergencystretch=\maxdimen
\usepackage[left,displaymath, mathlines]{lineno}
 
\DeclareMathAlphabet{\mathpzc}{OT1}{pzc}{m}{it}
\newcommand{\recl}[2]{${#1}< Re^{\text{c}} < {#2}$}     
\newcommand{\recu}[2]{${#1}< Re_{\text{c}}< {#2}$} 
\def\fig{Figure~}
\def\figs{Figures~}
\def\eqn{Eq.~}
\def\eqns{Eqs.~}
\def\tab{Table~}

\def\tsc#1{\csdef{#1}{\textsc{\lowercase{#1}}\xspace}}
\tsc{WGM}
\tsc{QE}
\tsc{EP}
\tsc{PMS}
\tsc{BEC}
\tsc{DE}
\newcommand{\rev}[1]{\textcolor{black}{#1}}     

\begin{document}
%
%
%
%
%
\begin{frontmatter} 
%
%
%
%
\title{{Critical parameters for \rev{non-Newtonian shear-thickening} power-law fluids flow across a channel confined circular cylinder}}
\author[labela,labelb]{Garima Vishal}
\author[labela,labelb]{Jyoti Tomar}
\author[labela,labelb]{Ram Prakash Bharti\corref{coradd}}
\ead{rpbharti@iitr.ac.in}
\address[labela]{Complex Fluid Dynamics and Microfluidics (CFDM) Lab, Department of Chemical Engineering, Indian Institute of Technology Roorkee, Roorkee - 247667, Uttarakhand, INDIA}
%
%
\cortext[coradd]{\textit{Corresponding author.} $^1${First author(s) contributed equally.}}
%
\begin{abstract}
\fontsize{12}{16pt}\selectfont
\noindent 
In this work, the critical parameters for an incompressible flow of non-Newtonian \rev{shear-thickening} power-law fluids across a channel confined circular cylinder have been investigated numerically. The governing equations have been solved by using the finite volume method for the wide range of power-law  ($1\le n\le 1.8$) fluids and for two values of wall blockage ratio ($\beta=2$ and 4). The present methodology has extensively been validated with numerical and experimental results available for limited conditions. \rev{Transitional insights of channel confined cylinder}, in particular, critical parameters indicating the transitions from creeping to \rev{separating flows (i.e., onset of} steady symmetric wake formation\rev{)}, and from steady symmetric wake to unsteady asymmetric wake formation \rev{(i.e., onset of vortex formation)} are investigated and presented in terms of the critical Reynolds numbers ($Re^{\text{c}}$ and $Re_{\text{c}}$).  The relative impacts of unconfined and confined flows on these critical parameters have also been explored.  In general, both \rev{onsets of the} flow separation and \rev{wake} asymmetry delayed with an increasing values of the power-law index ($n$) and the wall confinement ($\lambda$).   \rev{The dependence of critical $Re$ on $n$ for the confined (finite $\beta$) flow are, however, completely opposite to that for unconfined ($\beta=\infty$) flow, i.e., critical $Re$ decreased with increasing $n$. The influence of power-law index on the onset of vortex is quite stronger to that on the onset of wake formation. For instance, $Re^{\text{c}}$ for $\beta=(2, 4, \infty)$ altered from (12.5, 7.25, 6.25) to (30.5, 9.25, 0.75)  and the corresponding changes with $Re_{\text{c}}$ are noted from (84.5, 70.25, 46.5) to (449.5, 179.5, 33.5) as $n$ varied from 1 to 1.8, respectively.} The Stokes paradox \rev{(i.e., no creeping flow even as $Re\rightarrow 0$)} apparent with unconfined flow of power-law fluids is irrelevant in confined flows, under otherwise identical conditions.  Finally, the predictive correlations for critical $Re$ as a function of dimensionless parameters ($n$ and $\beta$) are presented for their easy use in engineering analysis.
\end{abstract}
\end{frontmatter}
%
%
\noindent 
\nomenclature[g0]{\textit{Greek letters}}{}
\nomenclature[d0]{\textit{Dimensionless groups}}{}
 \nomenclature[z0]{\textit{Abbreviations}}{}
%
 \nomenclature[zpiv]{PIV}{particle image velocimetry}
 \nomenclature[zfdm]{FDM}{finite difference method}
 \nomenclature[zfem]{FEM}{finite element method}
 \nomenclature[zfvm]{FVM}{finite volume method}
 \nomenclature[zgamg]{GAMG}{geometric-algebraic multi-grid}
 \nomenclature[zpiso]{PISO}{pressure-implicit split operator}
 \nomenclature[zopen]{OpenFOAM}{open source field operation and manipulation}
 \nomenclature[zquick]{QUICK}{quadratic upstream interpolation for convective kinematics}
\nomenclature[aCD]{$C_{\text{D}}$}{total drag coefficient (\eqn\ref{eq15}), dimensionless}
\nomenclature[aCDP]{$C_{\text{DP}}$}{pressure component of total drag coefficient, dimensionless}
\nomenclature[aCL]{$C_{\text{L}}$}{total lift coefficient (\eqn\ref{eq16}), dimensionless}
\nomenclature[aCDF]{$C_{\text{DF}}$}{frictional component of total drag coefficient, dimensionless}
\nomenclature[aCLP]{$C_{\text{LP}}$}{pressure component of total lift coefficient, dimensionless}
\nomenclature[aCLF]{$C_{\text{LF}}$}{frictional component of total lift coefficient, dimensionless}
\nomenclature[aD]{$D$}{diameter of a circular cylinder, m}
\nomenclature[aDt]{$\mathbf{D}$}{rate of strain tensor (\eqn\ref{eq5}), s$^{-1}$}
\nomenclature[af]{$f$}{body force, N}
\nomenclature[afv]{$f_{\text{v}}$}{frequency of vortex shedding, s$^{-1}$}
\nomenclature[aFD]{$F_{\text{D}}$}{total drag force per unit length of the cylinder, N/m}
\nomenclature[aFL]{$F_{\text{L}}$}{total lift force per unit length of the cylinder, N/m}
\nomenclature[aFDP]{$F_{\text{DP}}$}{pressure drag force per unit length of the cylinder, N/m}
\nomenclature[aFDF]{$F_{\text{DF}}$}{frictional drag force per unit length of the cylinder, N/m}
\nomenclature[aFLP]{$F_{\text{LP}}$}{pressure lift force per unit length of the cylinder, N/m}
\nomenclature[aFLF]{$F_{\text{LF}}$}{frictional lift force per unit length of the cylinder, N/m}
\nomenclature[aH]{$H$}{height of the computational domain, m}
\nomenclature[aII]{$I_{2}$}{second invariant of the strain rate tensor (\eqn\ref{eq7}), s$^{-2}$}
\nomenclature[aLuu]{$L_{\text{u}}$}{upstream length, m} 
\nomenclature[aLdd]{$L_{\text{d}}$}{downstream length, m}
\nomenclature[aL]{$L$}{length of the computational domain, m}
\nomenclature[am]{$m$}{fluid consistency index, Pa.s$^n$} 
\nomenclature[an]{$n$}{flow behavior index, dimensionless} 
\nomenclature[aP]{$p$}{pressure, Pa} 
\nomenclature[aRel]{$Re^{\text{c}}$}{lower critical $Re$ at onset of wake formation, dimensionless}
\nomenclature[aReu]{$Re_{\text{c}}$}{upper critical $Re$ at onset of wake asmmetry, dimensionless}
\nomenclature[aU]{$\mathbf{u}$}{velocity vector, m/s}
\nomenclature[aUa]{$u_{\text{avg}}$}{average velocity of the fluid at the inlet (\eqn\ref{eq11}), m/s}
\nomenclature[aUm]{$u_{\text{max}}$}{maximum velocity of the fluid at the inlet (\eqn\ref{eq11}), m/s}
\nomenclature[aux]{$u_{\text{x}}$}{x-component of the velocity vector, m/s}
\nomenclature[auy]{$u_{\text{y}}$}{y-component of the velocity vector, m/s}
\nomenclature[ax]{$x$}{stream-wise coordinate}
\nomenclature[ay]{$y$}{transverse coordinate}
\nomenclature[axo]{$\text{X}$}{critical $Re$ normalized w.r.t. corresponding unconfined flow (\eqns\ref{nrec}, \ref{nrec1}), dimensionless}
\nomenclature[ayo]{$\text{Y}$}{critical $Re$ normalized w.r.t. corresponding unconfined Newtonian flow (\eqns\ref{nrec}, \ref{nrec1}), dimensionless}
%
\nomenclature[gbeta]{$\beta$}{wall blockage ratio, dimensionless}
\nomenclature[glambda]{$\lambda$}{wall confinement ratio ($=\beta^{-1}$), dimensionless}
\nomenclature[geta]{$\eta$}{viscosity, Pa.s}
\nomenclature[grho]{$\rho$}{density of fluid, kg/m$^3$}
\nomenclature[gsigma]{$\sigma$}{total stress tensor, N/m$^2$}
\nomenclature[gtau]{$\tau$}{extra stress tensor,  N/m$^2$}
\nomenclature[dRey]{$Re$}{Reynolds number (\eqn\ref{eq14}), dimensionless}
\nomenclature[dStr]{$\mathit{St}$}{Strouhal number (\eqn\ref{eq17}), dimensionless}
\fontsize{10}{8pt}\selectfont
\renewcommand{\nompreamble}{\vspace{1em}}
\vspace{-2em}\rev{\printnomenclature[5em]}
%
%
\spacing{1.6}
%
%
\fontsize{12}{15pt}\selectfont
%
\section{Introduction}
%
\noindent 
Flow past cylinders of the circular and non-circular cross-sections is a dynamic area for research because of their fundamental and practical applications \citep[e.g., see][etc.]{Coutanceau1991,Eckelmann93,Williamson1996,Zdravkovich1997Book,Zdravkovich2003Book,Chhabra2006book,Chhabra2011,Michaelides2006book}. 
A reliable source of knowledge is therefore required in order to understand the hydrodynamic forces acting on the cylinder causing changes in the surrounding flow patterns. These phenomena can be observed in various aerodynamics, chemical, and process industries where cylindrical geometry is used for the thermal processing of materials. 
Further, sensors and probes are used to measure the flow rate and other parameters in the flowing fluid. 
\rev{For a Newtonian fluid flow over a cylinder, \citet{Zdravkovich2003Book} summarized that the blockage effects are negligible for the smaller confinement ($\lambda < 0.1$). The flow gets modified in the range of $0.1\le \lambda \le 0.6$, and suitable corrections can be made. The noticeable alteration of flow features beyond $\lambda > 0.6$ cannot be corrected based on available data. It is, however, not applicable at the very low Reynolds number ($Re$) in the two-dimensional laminar flow. The wall blockage effects are significant at low $Re$ even for negligible confinement ($\lambda < 0.001$). As briefed elsewhere \citep{Bharti2007a,Bharti2007b}, reliable information is broadly available on the Newtonian fluid flow across a channel confined cylinder.} 
\\\noindent 
\rev{Furthermore,} the wide ranging applications \rev{of the cylindrical geometry}  
\citep{Coutanceau1977a,Coutanceau1977b,Townsend1980,Zovatto2001,Chhabra2011} encounter both the Newtonian and non-Newtonian fluids. 
An extensive knowledge is required to handle the non-Newtonian fluids \citep{Chhabra2008Book,Malkin2012,Irgens2014} such as polymer 
solutions, lubricants, \rev{cosmetics, quicksands, asphalts, paints, pastes, creams, slurries, muds, sludge,} etc. experienced in processes and industries. One of the recent and greatest use of 
non-Newtonian shear-thickening \rev{(or dilatant)} power-law fluid, \rev{whose viscosity increases with increasing shear-rate}, can be seen in army as body armor or bulletproof jacket material  
\citep{Hanlon2006,Siuru2006,Boyle2010,Atherton2015,Matthews2016}. 
\\\noindent
While significant amount of literature is available on the flow of non-Newtonian fluids across a circular cylinder in both confined and unconfined 
arrangements \citep[e.g., see][etc.]{DAlessio1996,Chakraborty2004,Chhabra2004,DAlessio2004,Bharti2006,Sivakumar2006,Bhartit06,Patnana2009,Patnana2010, 
Chhabra2011,Bijjam2012,Al-Muslimawi13,Xiong2013,Tian2014,vishal15,Norouzi2015}, transitional insights of channel confined circular cylinder \rev{submerged in the non-Newtonian fluids} are still unknown. 
Therefore, this work aims to investigate the critical parameters (in particular, critical Reynolds numbers) for transitions from creeping 
to separating flows (i.e., onset of wake \rev{formation}), and from separating to transient flows (i.e., onset of \rev{wake instability or} vortex \rev{formation}). 
%
\section{Background literature}
%
\noindent 
Fluid flow over a circular cylinder in both confined and unconfined arrangements has been explored continuously over the decades 
\citep[e.g., see][etc.]{White1946,Takaisi1955,Coutanceau1977a,Coutanceau1977b,Townsend1980,Carte1995,Chen1995, 
Huang1995,zhao1999,zhao2000,Gupta2003,Khan2004,Mittal2006,Kumar2006, De2007,Cao2010,Sahu2010,Singha2010, Kanaris2011,Gautier13,Bayraktar2014,Kumar2014,Zhao2016,Thakur2018,Laidoudi17,Laidoudi18,Laidoudi_B18,Kumar2018,Zhang2019,Laidoudi20,Yasir2020,Laidoudi21}. The detailed and reliable information of hydrodynamic and heat transfer features of such flows have been reported 
in excellent review articles and books  
\citep[e.g., see][etc.]{Coutanceau1991,Williamson1996,Zdravkovich1997Book,Zdravkovich2003Book,Chhabra2006book,Chhabra2011,Michaelides2006book}.
Since the detailed literature of unconfined flow over a cylinder has been summarized in recent studies 
\citep{Bharti2006,Sivakumar2006,Patnana2009,Patnana2010,Pravesh2019ijhmt}, only relevant studies are mentioned herein.
For instance, the flow of a viscoelastic fluid based on an implicit four constant Oldroyd model has been investigated \citep{Townsend1980} by considering an infinite domain with a moving cylinder placed between the walls. For a Newtonian fluid flow at Reynolds number $Re=40$, 
the drag coefficient value was reported as 1.2. It was also shown that the low rotational speed has great significance in the case of a Newtonian fluid. Both drag and lift coefficients increase with an increase in rotational speed. Whereas an opposite behavior was seen for shear-thinning fluids, i.e., the drag tends to decrease with an increase in rotational speed. 
\citet{DAlessio1996} have used the first-order accurate finite difference method (FDM) to solve the stream function and vorticity formulation for an unconfined steady flow of power-law fluid across a cylinder. They presented the flow characteristics like drag coefficient, flow separation angle, wake length, and critical Reynolds number, etc. for limited flow conditions: $Re=5~ 
(0.65\le n\le 1.2)$, $Re=20~(0.8\le n\le 1.15)$ and $Re=40~(0.95\le n\le 1.1)$. Their results suggested the complex dependence of flow separation on power-law 
index ($n$), i.e., the critical Reynolds number was obtained to be $\sim 5$ and $\sim 6$ for $n = 1.2$ (shear-thickening) and 1 (Newtonian), 
respectively. Their drag values, unfortunately, appears to be in error \citep{DAlessio2004} due to unintended exclusion of a factor in one of their equation during post-processing of results. \citet{Chhabra2004} have replicated the work of \citet{DAlessio1996} by using the corrected equation and 
second-order accurate FDM for $1\le Re\le 40$ and $0.2\le n\le 1.4$.  This flow field \citep{Chhabra2004}  was used by  \citet{Soares2005}
to explore the forced convection heat transfer characteristics of power-law fluids across an unconfined cylinder. Subsequently, a detailed systematic parametric study \citep{Bharti2006} of an unconfined steady flow of power-law fluids across a cylinder was performed by using the finite volume method (FVM) for $5\le Re\le 40$ and $0.6\le n\le 2$. These investigations have qualitatively as well quantitatively suggested the stronger dependence of transitional behavior of flow separation, wake and vortex formations on the fluid rheological behavior. 
The flow transitional regimes, however, have not been systematically demarcated, except for a couple of flow conditions.  
\citet{Sivakumar2006} focused on the investigation of the critical parameters for non-Newtonian power-law fluids flow across an unconfined circular cylinder. They reported the critical values of the Reynolds number ($Re_{\text{c}}$ and $Re^{\text{c}}$) as a function of the power-law index ($0.3\le 
n\le 1.8$) for the onset of wake separation and the onset of transition from steady symmetric to steady asymmetric wake formation. 
The wake separation was seen to postpone from $Re_{\text{c}}=6.5$ to 12 as the fluid behaviour changed from Newtonian ($n=1$) to shear-thinning ($n=0.3$), 
whereas it prepones from $Re_{\text{c}}=6.5$ to 1 as the fluid behaviour changed from Newtonian ($n=1$) to shear-thickening ($n=1.8$).
Similarly, they noted that in case of shear-thinning fluid, with an increase in $n<1$, the transition from steady wake to unsteady wake delays 
(critical $Re$ shifts to a higher value), whereas in case of shear-thickening fluid, the transition preponed with increase in $n>1$. The critical $Re$ values further suggested an appearance of `Stokes paradox' \citep{Tanner1993,MPaloka2001} for the power-law fluids flow over an unconfined cylinder. 
 These stronger dependencies of flow regimes on fluid rheology motivated us to explore the transitional behavior of regimes for the flow of non-Newtonian power-law fluids across a channel confined cylinder.
\\
Fluid flow across a channel confined cylinder has been investigated by various researchers over the decades \citep[e.g., 
see][etc.]{Chen1995,zhao1999,zhao2000,Zovatto2001,Gupta2003,Chakraborty2004,Khan2004,Sahin2004,Bharti2007a,Bharti2007b,Rehimi2008,Bijjam2012,Zhao2016,
Mathupriya2018}. Since the detailed literature on confined flow over a cylinder has been briefed elsewhere 
\citep{Bharti2007a,Bharti2007b}, only relevant studies are mentioned herein. 
For instance, \citet{Zovatto2001} explored the flow characteristics of Newtonian fluid over a cylinder confined in a channel by using the finite element method (FEM). They observed delay in the flow transition (from symmetric wake to periodic vortex shedding) with an increase in wall confinement. Vorticity contours were also reported for a steady state regime and observed that when the cylinder was placed in the middle of the two 
walls, wake was symmetric but as the cylinder shifted towards one of the walls, a significant reduction in wake vorticity was observed. 
\citet{Sahin2004} have analyzed the wall effects in the two-dimensional flow past a circular cylinder using the finite volume method (FVM). 
Critical Reynolds number and Strouhal number was calculated for different wall confinements ($0.1\le \lambda\le 0.9$). 
For $\lambda=0.5$, the critical Reynolds number was reported as 125.23. A monotonic increase in critical Reynolds number, as well as Strouhal number, was observed with an increase in blockage ratio. 
Further, \citet{Bharti2007a,Bharti2007b} have explored the two-dimensional Poiseuille flow of non-Newtonian power-law fluids across a channel confined circular cylinder using the finite volume method (FVM). Their parametric studies have reported both detailed as well as local flow and forced convection characteristics by systematic variations of wide ranges of flow governing and influencing parameters ($1.1\le \beta\le 4$, $1\le Re\le 40$, $0,2\le 
n\le 1.8$ and $1\le Pr\le 100$). The dependence of wake structure in Newtonian fluids on wall confinement appears to be consistent with other studies \citep{Carte1995,Sahin2004,Rehimi2008}. The wake size was observed to enhance with decreasing value of the flow behavior index ($n$). 
Because of the wall confinement effects, the flow separation found to postpone (or prepone) in shear-thickening (or shear-thinning) fluids.
\citet{Rehimi2008} have conducted 2D-2C-PIV experiments to investigate the confined ($\beta=3$) flow downstream of a circular cylinder placed between parallel walls for $30\le Re\le 277$. Their results compared well with the theoretical solutions  \citep{Lundgren1964} based on fourth order Range-Kutta method to calculate pathlines, and bilinear interpolation to find particle velocity. 
The first instability appeared at critical Reynolds number $Re^c$=108 was in good match with the simulation results, i.e., $Re^c$=97.5 \citep{Carte1995} and $Re^c$=101 \citep{Sahin2004}, respectively. 
They also found that the size of the recirculation region was greater as compared to that in an unconfined flow configuration. This effect can be argued on the basis that the wall effects stabilize and flatter the mean recirculation region in the case of confined flow \citep{Carte1995,Sahin2004,Bharti2007a}. Subsequently, 
\citet{Bijjam2012} have explored two-dimensional unsteady flow characteristics of power-law fluids across a channel confined cylinder for  $50\le Re\le 100$ and $0.4\le n\le 1.8$ at $\beta=4$. They reported smooth wake formation at $Re=50$ for $0.4\le n\le 
1.8$  and the size of symmetric vortices decreased with increasing $n$.  At $Re=75$, unsteady flow for $0.4\le n\le 1.2$ and steady flow for $1.2\le 
n\le 1.8$ is reported due to the higher damped nature of effective viscosity of shear-thickening ($n>1$) fluid. Similarly, the flow was recorded to be unsteady 
for $0.4\le n\le 1.4$ and steady for $1.4\le n\le 1.8$ at $Re=100$.
Further, \citet{kumar2016} have investigated an onset of vortex shedding and the effects of Reynolds and Prandtl numbers for confined flow over a semi-circular cylinder. For $\lambda=0.25$, the onset of vortex shedding is noticed at $ Re = 69.5\pm0.5$ for a Newtonian fluid. 
\\
The in-depth analysis of existing literature on the flow over a channel confined circular cylinder suggests that the critical parameters for Newtonian fluid flow  are known for very limiting governing and influencing parameters. To the best of our knowledge, none of the prior studies has revealed the detailed characterization of confined flow regimes for non-Newtonian fluids. The corresponding features for unconfined cylinder, however, have been established in the literature \citep{Sivakumar2006}. The present work, therefore, aims to strengthen the existing literature through numerical investigation of critical parameters indicating the onset of wake formation and the onset of wake instability for the flow of non-Newtonian power-law fluids over a channel confined circular cylinder by systematic variation of the Reynolds number ($Re$) for a broader range of wall blockage ratio ($\beta$) and flow behavior index ($n$).
%
\section{Problem statement}
\begin{figure}[b]
\centering
\includegraphics[width=1\linewidth]{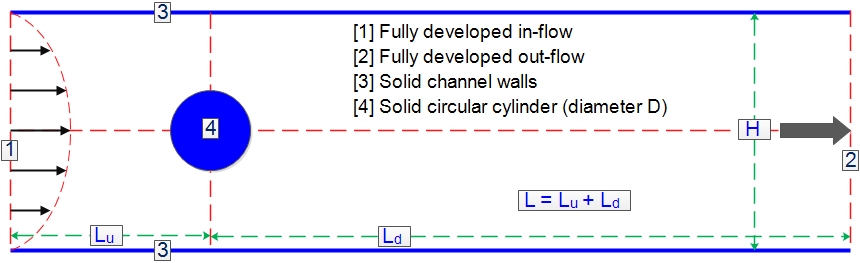}
\caption{Schematic representation of flow across a channel confined circular cylinder with physical boundary conditions.}
\label{fig:1}
\end{figure}

%
%
\noindent 
Consider a two-dimensional (2-D) fully developed flow  over an infinitely long circular cylinder (diameter $D$)   confined between the middle ($H/2$) of the two parallel plane walls separated by distance $H$ (\fig\ref{fig:1}). 
The wall blockage ratio ($\beta$) is defined as  $\beta=H/D$ \rev{and the wall confinement ratio ($\lambda$) is given as $\lambda=\beta^{-1}$}.   The flow of incompressible non-Newtonian power-law fluid is approaching a cylinder placed at upstream length ($L_{\text{u}}$) measured from the inlet to center of the cylinder, and the outlet (or exit) boundary is located at downstream length ($L_{\text{d}}$)  from the center of a cylinder. 
The total length and height of computational domain are $L~(= L_{\text{u}} + L_{\text{d}})$ and $H$, respectively. 
%
\\
Based on the above approximations, the flow governing equations, namely, mass continuity and momentum transport equations, can be written as follow.
\begin{eqnarray}
\nabla\cdot \mathbf{u} = 0 \label{eq1}\\
\rho\left(\frac{\partial \mathbf{u}}{\partial t} + \mathbf{u}\cdot\nabla \mathbf{u} - f\right) - \nabla\cdot\mathbf{\sigma} = 0 \label{eq2}
\end{eqnarray}
where $\rho$, $\mathbf{u}$,  $f$ and $\mathbf{\sigma}$  denote for the fluid density, velocity vector, the body force and the stress tensor, respectively. The stress 
tensor ($\sigma$), the summation of the isotropic pressure ($p$) and deviotoric stress tensor ($\tau$), is given by \eqn(\ref{eq3}).
\begin{equation}
\mathbf{\sigma} = -pI+\mathbf{\tau} \label{eq3}
\end{equation}
The rheological equation of state for incompressible fluids is given elsewhere \citep{Bird2006Book,Chhabra2008Book,Mory2011Book,Darby2017Book} as 
follows.
\begin{equation}
\mathbf{\tau} = 2\eta\mathbf{D}\label{eq4}
\end{equation}
The rate of strain tensor ($\mathbf{D}$) is given by
\begin{equation}
\mathbf{D} = \frac{1}{2}\left[\Delta \mathbf{u} + (\Delta \mathbf{u})^T\right] \label{eq5}
\end{equation}
The second invariant ($I_2$) of the rate of strain tensor ($\mathbf{D}$) for two-dimensional flow is given by
\begin{equation}
I_2=2(\mathbf{D}:\mathbf{D}) 
\label{eq7}
\end{equation}
For a power-law fluid, the apparent viscosity ($\eta$) is given by
\begin{equation}
\eta= m \dot{\gamma}^{(n-1)}\qquad\text{where}\qquad \dot{\gamma} = \sqrt{I_2/2} \label{eq6}
\end{equation}
where $m$ and $n$ being the power-law fluid consistency index and the flow behaviour index of the fluid ($n < 1$, $=1$, $> 1$ correspond to a  shear-thinning, a Newtonian and a shear-thickening fluid). 
\rev{ The fluid consistency index ($m$) represents for the shear-independent average viscosity of the fluid, whereas, the flow behaviour index ($n$) determines the extent of deviation of fluid behaviour from Newtonian nature.
The apparent viscosity ($\eta$) of shear-thickening ($n>1$) fluids decreases and shear-thinning ($n<1$) fluids increases with decreasing shear rate ($\dot{\gamma}$).} 
\\\noindent
The flow problem under consideration (\fig\ref{fig:1}) is subjected to the following physically realistic boundary conditions. 
The flow is assumed to be fully developed at the inlet ($x=0$), i.e., left boundary. Mathematically, the following conditions are applied at the inlet: 
\begin{equation}
u_{\text{x}} =u_{\text{p}}(y,n) \qquad\mbox{and}\qquad u_{\text{y}} = 0
\label{eq9}
\end{equation}
where, the fully developed velocity profile for the laminar flow of power-law fluids through a channel (of height $H$) is given by
\citet{Bharti2007a,Bharti2007b} as follows.
\begin{equation}
u_{\text{p}}(y,n) = \left[1-\left| 1-\frac{2y}{H}\right|^{(n+1)/n} \right]{u}_{\text{max}} 
 \qquad\mbox{for}\qquad 0\le y\le H
\label{eq10}
\end{equation}
The maximum velocity (${u}_{\text{max}}$) is related to the area-averaged velocity (${u}_{\text{avg}}$) as follows.
\begin{equation}
{u}_{\text{max}} =\left(\frac{2n+1}{n+1}\right){u}_{\text{avg}}
\label{eq11}
\end{equation}
The standard no-slip condition has been applied at the lower ($y=0$) and upper ($y=H$) channel walls, and on the surface of the 
cylinder, i.e.,
\begin{equation}
u_{\text{x}} =0 \qquad\mbox{and}\qquad u_{\text{y}} = 0\label{eq12}
\end{equation}
The Neumann condition has been imposed on the exit ($x=L$) boundary as follows.
\begin{equation}
\frac{\partial u_{\text{x}}}{\partial x} =0 \qquad\mbox{and}\qquad \frac{\partial u_{\text{y}}}{\partial x} =0\label{eq13}
\end{equation}
The computations are performed in the full computational domain shown in  \fig\ref{fig:1}. 
The numerical solution of the above mentioned governing equations (\eqns\ref{eq1} and \ref{eq2})
in conjunction with boundary conditions (\eqns\ref{eq9} - \ref{eq13}) results in the velocity ($\mathbf{u}$) and pressure ($p$) fields. 
\\\noindent
At this point, it is important to introduce some definitions used in this work. The dimensionless parameters are obtained by using $D$, ${u}_{\text{avg}}$, $D/{u}_{\text{avg}}$,  $\rho {u}_{\text{avg}}^2$, $({u}_{\text{avg}}/D)^2$, $m({u}_{\text{avg}}/D)^n$,  $m({u}_{\text{avg}}/D)^{n-1}$ as the scaling variables for length, velocity, time, pressure, rate of strain, shear stress,  and viscosity, respectively. 
\\\noindent
\textit{Reynolds number} ($Re$)  for non-Newtonian power-law fluid flow is defined as follow:
\begin{equation}
Re = \frac{\rho D^n {u}_{\text{avg}}^{2-n}}{m} \label{eq14}
\end{equation}
\rev{The effect of fluid consistency index ($m$) can be accounted through variation of $Re$ for a given density of fluid ($\rho$), characteristic length ($D$) and characteristic velocity (${u}_{\text{avg}}$).}
\\\noindent
The \textit{lower critical Reynolds number} ($Re^c$), and \textit{upper critical Reynolds number} ($Re_c$) are defined as the Reynolds numbers at which the flow transits from creeping to separating (i.e., onset of wake formation), and the flow experiences a transition from the two-dimensional (2-D) steady `symmetric' flow to `asymmetric' flow, i.e., onset of vortex formation, as defined elsewhere \citep{Sivakumar2006}. 
The 2-D symmetric wake flow regime occurs for the Reynolds number range in between these two critical limits ($Re^c \le Re \le Re_c$). 
\\\noindent
The \textit{total drag coefficient} ($C_{\text{D}}$)  can be defined as the sum of the pressure and frictional components of drag  as follows.
\begin{equation}
C_{\text{D}}=C_{\text{DP}}+C_{\text{DF}} \qquad \Rightarrow\qquad \frac{F_{\text{D}}}{(1/2)\rho {u}_{\text{max}}^2D} = \frac{F_{\text{DP}}}{(1/2)\rho {u}_{\text{max}}^2D} + \frac{F_{\text{DF}}}{(1/2)\rho {u}U_{\text{max}}^2D} 
\label{eq15}
\end{equation}
where $F_{\text{D}}$ is the total drag force per unit length of cylinder.  \rev{The $C_{DP}$  and $C_{DF}$  are the pressure and frictional contributions of  $C_{\text{D}}$.} The $F_{\text{DP}}$ and $F_{\text{DF}}$ are the pressure and frictional contributions of $F_{\text{D}}$, as defined elsewhere \citep{Sivakumar2006,Bharti2007a,Bharti2007b,Patnana2009,Patnana2010}. 
\\\noindent
The \textit{total lift coefficient} ($C_{\text{L}}$) can be defined as the sum of the pressure and frictional lift coefficients as follows.
\begin{equation}
C_{\text{L}}=C_{\text{LP}}+C_{\text{LF}} \Rightarrow \frac{F_{\text{L}}}{(1/2)\rho {u}_{\text{max}}^2D} = \frac{F_{\text{LP}}}{(1/2)\rho {u}_{\text{max}}^2D} + \frac{F_{\text{LF}}}{(1/2)\rho {u}_{\text{max}}^2D} 
\label{eq16}
\end{equation}
where $F_{\text{L}}$ is the total lift force per unit length of cylinder. \rev{The $C_{LP}$  and $C_{LF}$  are the pressure and frictional contributions of  $C_{\text{L}}$.}  The $F_{\text{LP}}$ and $F_{\text{LF}}$ are the pressure and frictional contribution of $F_{\text{L}}$, as defined elsewhere \citep{Sivakumar2006,Patnana2009,Patnana2010}. 
\\\noindent
The \textit{Strouhal number} ($\mathit{St}$), the dimensionless frequency of vortex shedding, is defined as
\begin{equation}
\mathit{St} =\frac{f_{\text{v}}D}{{u}_{\text{avg}}}
\label{eq17}
\end{equation}
Here, $f_{\text{v}}$ is the frequency of vortex shedding. The \textit{critical Strouhal number}  ($\mathit{St}_{\text{c}}$) is defined as the Strouhal number ($\mathit{St}$) at 
the {upper critical Reynolds number} ($Re_{\text{c}}$). 
For a steady flow regime, the values of both the lift coefficient ($C_{\text{L}}$) and the Strouhal number ($\mathit{St}_{\text{c}}$) tends to zero.
%
\\\noindent 
\rev{The above detailed mathematical model has been solved by using the unstructured finite volume method (FVM). The subsequent section has briefly discussed the numerical methodology.}
%
\section{\rev{Numerical method}}
%
\noindent 
\rev{
In this work, the flow field equations in conjunction with realistic boundary conditions have been solved using the unstructured finite volume method (FVM). 
Since the detailed discussion of the finite volume method (FVM) is documented in various standard text/reference books \citep[e.g., see ][etc.]{Anderson95,Blazek01,Versteeg2011,Barth17,Ferziger20,sharma2021}, only the brief approach is recapitulated here. 
In the finite volume (FV) approach,  the general transport equation, i.e., governing partial differential equations, for a general scalar variable ($\phi$) are first integrated over the finite control volumes (CVs) into which the domain has been discretized \citep{Versteeg2011,sharma2021}. 
\begin{equation}
	\int_{\Omega} \underbrace{\frac{\partial (\rho\phi)}{\partial t}dV}_{\text{transient term}}
+	\int_{\Omega} \underbrace{\nabla\cdot(\rho\mathbf{u}\phi)dV}_{\text{convective term}}
-	\int_{\Omega} \underbrace{\nabla\cdot(\rho\Gamma_{\phi}\nabla \phi)dV}_{\text{diffusive term}}
=	\int_{\Omega} \underbrace{S_{\phi}(\phi)dV}_{\text{source term}}
	\label{gte}
\end{equation}
The Gauss theorem ($\int_{\Omega} (\nabla\cdot \mathbf{u})dV = \oint_{\partial\Omega} d\mathbf{S}\cdot\mathbf{u}$) is subsequently applied to transform the volume integral of the convection and diffusion terms into surface integral. 
Here $\oint_{\partial\Omega}$ is the surface integral over the control surface $\partial\Omega$. The surface integrals are further linearized by interpolating the cell centered values to the face centers of CV. The discrete equations for each term yielded as follows. 
\begin{eqnarray}
	\oint_{\partial\Omega} \underbrace{d\mathbf{S}\cdot(\rho\mathbf{u}\phi)}_{\text{convective term}} 
	&=& \sum_{f} \left[\int_{f} d\mathbf{S}\cdot(\rho\mathbf{u}\phi)\right]
	\approx \mathbf{S}_{f}\cdot(\overline{\rho\mathbf{u}\phi})_{f}
	= \mathbf{S}_{f}\cdot({\rho\mathbf{u}\phi})_{f}
	\label{gconv} \\
%
	\oint_{\partial\Omega} \underbrace{d\mathbf{S}\cdot(\rho\Gamma_{\phi}\nabla \phi)} _{\text{diffusive term}}
	&=& \sum_{f} \left[\int_{f} {d\mathbf{S}\cdot(\rho\Gamma_{\phi}\nabla \phi)} \right]
	\approx \mathbf{S}_{f}\cdot(\overline{\rho\Gamma_{\phi}\nabla \phi})_{f}
	= \mathbf{S}_{f}\cdot(\rho\Gamma_{\phi}\nabla \phi)_{f}
	\label{gcond}
	\\
	\underbrace{(\nabla \phi)_P}_{\text{gradient term}}
	 &=& \frac{1}{\Omega} \sum_{f} (\mathbf{S}_{f}\phi_{f})
	\label{ggrad}
\\
	\int_{\Omega} \underbrace{S_{\phi}(\phi)dV}_{\text{source term}}
	&=&  S_cV_P + S_pV_P\phi_P
	\label{gsou}
\end{eqnarray}
The integrants in the above \eqns(\ref{gconv}) - (\ref{gsou}) are approximated by the second order accurate mid point rule. 
The centroid (P) gradients are approximated by the Gauss theorem, which is second order accurate.  The $d\mathbf{S}$ represents an infinitesimal surface element with associated normal ($\mathbf{n}$) pointing 
outwards of the surface $\partial\Omega$ and $\mathbf{n}d\mathbf{S}=d\mathbf{S}$. The source term approximation is exact for constant or linearly varying $S_{\phi}$ with in CV, otherwise second order accurate. In \eqn(\ref{gsou}), $S_c$ and $S_p$ are the constant (or linear) and non-linear parts of source term. 
\\\noindent 
By using the above approximations (\eqns\ref{gconv} -\ref{gsou}), the general transport equation (\eqn\ref{gte}) over all CVs can be written in the following semi-discrete form. 
\begin{equation}
	\int_{\Omega} \underbrace{\frac{\partial (\rho\phi)}{\partial t}dV}_{\text{transient term}}
+	\sum_{f} \underbrace{\mathbf{S}_{f}\cdot({\rho\mathbf{u}\phi})_{f}}_{\text{convective flux, } J_{c,f}}
-	\sum_{f} \underbrace{ \mathbf{S}_{f}\cdot(\rho\Gamma_{\phi}\nabla \phi)_{f}}_{\text{diffusive flux, } J_{d,f}}
=	\underbrace{(S_cV_P + S_pV_P\phi_P)}_{\text{source term}}
	\label{gted}
\end{equation}
The surface  fluxes are obtained at the faces of CV without integrating within CV. The conservativeness of FVM is retained through this transformation. Since all variables are computed and stored at the centroid (P) of CVs, face (f) values appearing in the convective and diffusive fluxes ($J_{c,f}$ and $J_{d,f}$) are computed by using the interpolation from the centroid values of CVs at both sides of face. In this work, the temporal derivative, convective and diffusive fluxes terms are discretized using time-implicit scheme, 3rd order accurate QUICK (Quadratic Upstream Interpolation for Convective Kinematics) scheme \citep{Leonard79,Hayase92}, and 2nd order accurate CD (central difference) scheme, respectively. The algebraic equations resulting from the above discussed procedure are solved using the solution procedure discussed in subsequent section.   
}
%
\section{Solution procedure}
%
\noindent 
In this work, the flow field equations in conjunction with realistic boundary conditions have been solved using the unstructured finite volume method (FVM) based open-source solver OpenFOAM \citep{openfoam,Jasak07openfoam,Moukalled16}.
\rev{The OpenFOAM solver uses a `collocated grid' approach on an unstructured polyhedral non-uniform grid with arbitrary grid elements.  In this grid arrangement \citep{Meier99}, all the flow variables are computed and stored on the `centroid' of a control volume (CV). Implicit approach is used to discretize the temporal derivative.} The \rev{sufficiently refined} suitable unstructured grid has been generated by using an open-source program. The ``{Non-Newtonian Icofoam}'' (transient solver for incompressible, laminar flow of non-Newtonian fluids) solver has been used to account for the rheological model behavior. The ``{generalized GAMG}'' (geometric-algebraic 
multi-grid) solver is used to solve the algebraic equations. The ``{smoothSolver}'' (solver using a smoother for both symmetric and asymmetric matrices) is used to obtain the velocity field. The ``{PISO}'' (pressure-implicit split-operator) scheme is utilized for coupling of pressure-velocity and non-Newtonian power-law model for viscosity.  Relative tolerance of $10^{-6}$ has been used in computations of velocity and pressure fields. 
%
\section{Choice of numerical parameters}
%
\noindent
The complex fluid flow problems have a significant concern about the reliability and accuracy of numerical results. Their hydrodynamic nature is intensely sensitive to relatively small changes in flow governing and influencing parameters. Therefore, a suitable choice of numerical parameters is vital to obtain the numerical results free from numerical artifacts, ends effects, etc.
The problem under consideration has the three flow governing parameters (namely, wall blockage $\beta$, Reynolds number $Re$ and flow behavior index $n$) and two flow influencing parameters (upstream and downstream lengths of the channel, $L_{\text{u}}$ and $L_{\text{d}}$; and grid points distribution). The correct choice of influencing parameters is obtained by performing the domain and grid independence tests over the range of flow governing parameters considered herein, to ensure that the new results presented hereafter are free from the \rev{numerical artifacts and} ends effects.
\subsection{Domain independence test}
\noindent
The domain independence study has been carried in two steps, (a) $L_{\text{d}}$ test with a fixed $L_{\text{u}}$, and (b) $L_{\text{u}}$ test with the selected $L_{\text{d}}$ in previous step. 
First, the downstream length ($L_{\text{d}}$) independence test has been performed by systematic variation of $\rev{L_{\text{d}}^{\ast}=} L_{\text{d}}/D$ as 20, 40, 60 and 80 with the fixed \rev{value of} upstream length ($\rev{L_{\text{u}}^{\ast} =} L_{\text{u}}/D=10$). \tab\ref{Tab:1} summarizes the influence of downstream length ($L_{\text{d}}^{\ast}$) on the drag and lift coefficients ($C_{\text{D}}$
and $C_{\text{L}}$) for the extreme values of the blockage ratio ($\beta=1.1$ and 4) and flow behavior index ($n=1$ and 1.8) at a fixed Reynolds number
($Re=40$).
The G2 grid (details shown in \tab\ref{Tab:2}) is used in the domain independence test cases.
While $C_{\text{D}}$ values have negligible variation with an increase in $L_{\text{d}}^{\ast}$, $C_{\text{L}}$ values show stronger dependence at lower $L_{\text{d}}^{\ast}$.
Keeping in mind the excessive enhancement in
computational efforts, \rev{i.e., simulation time,} with insignificant changes in the drag and lift values for $L_{\text{d}}^{\ast}>40$, the downstream length $L_{\text{d}}^{\ast}=40$ is believed to be
sufficient to produce the accurate results.
\begin{table}
\centering
\caption{Domain independence test for flow around a  channel confined cylinder.}
\label{Tab:1}
\fontsize{9}{12pt}\selectfont
\begin{tabular}{cccccc}
\hline
$Re=40$ &\multicolumn{2}{l}{$\beta=4$, $n=1$}&\multicolumn{2}{l}{$\beta=4$, $n=1.8$}&{$\beta=1.1$, $n=1.8$}\\ \cline{2-6}
&$C_D$&$10^4C_L$&$C_D$&$10^4C_L$&$C_D$\\  
\hline
$L_{\text{d}}^{\ast}$ &\multicolumn{5}{l}{(a) Downstream length ($L_{\text{d}}^{\ast}$) test with $L_{\text{u}}^{\ast}$=10}\\ 
\hline
20	&	1.706567	&	-1.64	&	2.582733	&	-5.80	&	43378.69	\\
40	&	1.706576	&	-2.72	&	2.582745	&	-9.60	&	43378.90	\\
60	&	1.706573	&	-3.80	&	2.582736	&	-9.63	&	43378.79	\\
80	&	1.706565	&	-3.89	&	2.582725	&	-9.63	&	43378.69	\\
\hline
$L_{\text{u}}^{\ast}$ &\multicolumn{5}{l}{(b) Upstream length ($L_{\text{u}}^{\ast}$) test with $L_{\text{d}}^{\ast}$=40}\\ 
\hline
10	&	1.706576	&	-2.72	&	2.582745	&	-9.62	&	43378.90	\\
15	&	1.706582	&	-3.05	&	2.599991	&	-9.66	&	43377.03	\\
20	&	1.706595	&	-3.33	&	2.608500	&	-9.74	&	43376.55	\\
\hline
\end{tabular} 
\end{table}
\\\noindent
Having selected the downstream length ($L_{\text{d}}^{\ast}=40$), the upstream length ($L_{\text{u}}^{\ast}$) is tested by variation of $L_{\text{u}}^{\ast}$ as 10, 15 and 20.
\tab\ref{Tab:1} also shows the influence of upstream length ($L_{\text{u}}^{\ast}$) on $C_{\text{D}}$ and $C_{\text{L}}$ for two extreme blockage ratio ($\beta=1.1$ and 4) and for the two
extreme values of flow behavior index ($n=1$ and 1.8).
The influence of $L_{\text{u}}^{\ast}$ is seen to be qualitatively similar to that observed in $L_{\text{d}}^{\ast}$ test. A negligible alteration in the drag and lift coefficients for $L_{\text{u}}^{\ast} >15$ is observed with an excessive increase in computational cost. Therefore, based on the trade-off between computational efforts and accuracy, upstream length $L_{\text{u}}^{\ast}=15$ and downstream length $L_{\text{d}}^{\ast}=40$ are believed to be sufficient to produce the results free from end effects.
\subsection{Grid independence test}
\noindent
The grid independence test is performed by taking various unstructured \rev{non-uniform} grids (G1 to G10) with different mesh sizes at the channel edges and the varying number of points over the \rev{circumference of a} cylinder. The grid specifications are noted in \tab\ref{Tab:2}.
Included in \tab\ref{Tab:2} is the dependence of grid structure on the drag and lift coefficients ($C_{\text{D}}$ and $C_{\text{L}}$) for two extreme values of the blockage ratio ($\beta=1.1$ and 4) and flow behavior index ($n=1$ and 1.8) at a fixed Reynolds number ($Re=40$).
\begin{table}
\centering
\caption{Grid independence test at $Re=40$ for blockage ratio of $\beta$=1.1 and 4. (Nc, $\delta$, and $\Delta$ are the number of grid points on the surface of cylinder, the minimum and maximum grid spacing, respectively.)}
\label{Tab:2}
\fontsize{9}{12pt}\selectfont
\begin{tabular}{ccccccccc}
\hline
\multicolumn{4}{l|}{Grid specifications}
&\multicolumn{2}{l}{$\beta=4$, $n=1$}&\multicolumn{2}{l}{$\beta=4$, $n=1.8$}&{$\beta=1.1$, $n=1.8$}\\ \hline
No.	&Nc	&$\delta^{-1}$	&	$\Delta^{-1}$	&	$C_D$	&	$10^4C_L$	&	$C_D$	
&	$10^4C_L$	&$C_D$	\\
\hline
G1	&240	&60		&60		&1.70649	&-0.27	&2.5858	&-2.00	&37177.30	\\
G2	&240	&100	&100	&1.70658	&-3.05	&2.6000	&-9.66	&43377.03	\\
G3	&240	&160	&160	&1.70686	&-0.90	&2.4146	&-9.55	&43496.43	\\
G4	&360	&60		&60		&1.71123	&-5.10	&2.5695	&-8.35	&43172.65	\\
G5	&360	&100	&60		&1.70569	&-7.80	&2.4586	&-7.00	&--		\\
G6	&360	&100	&100	&1.71102	&-7.70	&2.5846	&-17.00	&43544.74	\\\hline
G7	&480	&100	&60		&1.70502	&1.00	&2.4588	&-5.00	&--		\\\hline
G8	&480	&100	&100	&--	&--	&--	&--	&45144.74	\\
G9	&480	&160	&60		&1.70519	&2.40	&--	&--	&--		\\
G10	&600	&100	&100	&--	&--	&--	&--	&45740.47	\\\hline
\end{tabular}
\end{table}
An analysis of \tab\ref{Tab:2} shows the insignificant changes in $C_{\text{D}}$ and $C_{\text{L}}$ values with the refinement of the grid structure for Newtonian ($n=1$) fluids. However, the grid structure played a significant role at a more considerable value of the flow behavior index ($n=1.8$). Further, the computational efforts have enhanced many folds in obtaining the solutions by refining the grid structure from G1 to G10.
The strong non-linearities associated with complex fluid flow simulations require a sufficiently refined grid to capture the sharp changes in the gradients that may encounter during the computations. Overall analysis, thus, suggests the adequacy of grid G7 with reasonable computational efforts for the ranges of conditions being considered herein this work. 
Based on our previous experiences, grid G7 is believed to be sufficiently refined to produce the results to be reliable and accurate within $\pm 1-2\%$.
%
\section{Results and discussion}
%
\noindent 
In this work, 2-D \rev{transient} simulations \rev{for flow over a channel confined cylinder} have been performed for the channel blockage ratio of $\beta=4$ and 2 over the wide range of power-law index ($1\le n\le 1.8$). The Reynolds number ($Re$) is varied in the gaps of 0.5 and 1,  starting from critical $Re$ for unconfined ($\beta=\infty$) flow of power-law fluids, until the critical conditions are obtained. The critical parameters have been deduced through visualization of flow streamlines ($\psi$), pressure coefficient ($C_{\text{p}}$), friction coefficient ($C_{\text{f}}$),  and lift and drag coefficients ($C_{\text{L}}$ and $C_{\text{D}}$) profiles. 
\\ \noindent
Before the presentation of new results, the present numerics have been validated with the existing literature for its efficacy and reliability. 
\begin{table}
	\centering
	\caption{Comparison of drag coefficient values for steady power-law fluid flow over a cylinder.}
	\label{Tab:3}
	\fontsize{9}{12pt}\selectfont
	\begin{tabular}{ccccccccc}
		\hline
		&	\multicolumn{2}{l}{$\beta=\infty$}	&	\multicolumn{6}{l}{$\beta=4$}		\\\cline{2-9}
		&	$Re=40$	&	$Re=20$	&	\multicolumn{3}{l}{$Re = 40$}	&	\multicolumn{3}{l}{$Re = 1$}		\\\cline{2-9}
		Source	&	$n=1$	&	$n=1$	&	$n=1$	&	$n=1.2$	&	$n=1.8$	&	$n=1$	&	$n=1.2$	&	$n=1.8$	\\
		\hline
		\citet{Dennis1970}	&	1.5220	&	2.045	&	-	&	-	&	-	&	-	&	-	&	-	\\
		\citet{Fornberg1980}	&	1.4980	&	2.000	&	-	&	-	&	-	&	-	&	-	&	-	\\
		\citet{Park1998}	&	1.5100	&	2.010	&	-	&	-	&	-	&	-	&	-	&	-	\\
		\citet{Niu2003}&	1.5740	&	2.111	&	-	&	-	&	-	&	-	&	-	&	-	\\
		\citet{Bharti2007a}	&	-	&	-	&	1.7034	&	1.8793	&	2.4765	&	28.536	&	32.591	&	51.453	\\
		\citet{Bijjam2012}	&	-	&	-	&	1.7039	&	1.8781	&	2.4770	&	-	&	-	&	-		\\\hline
		Present work	&	1.5365	&	2.0547	&	1.7050	&	1.8730	&	2.4588	&	28.566	&	32.597	&	51.429	\\
		\hline
	\end{tabular}
\end{table}
\noindent \tab\ref{Tab:3} compares the present drag coefficient ($C_{\text{D}}$) values with the existing literature for Newtonian ($n=1$) and non-Newtonian  ($n=1.2$ and 1.8) 
fluids flow across a cylinder placed in confined ($\beta=4$) and unconfined ($\beta=\infty$) mediums for three values of Reynolds number ($Re=1$, 20 and 40).
It can clearly be seen that the present results are matching closely with the literature values. For instance, an analysis of \tab\ref{Tab:3} 
yields the maximum relative difference, $\delta_{\text{r}}(\phi)=\left|({\phi_{\text{literature}}-\phi_{\text{present}})}/{\phi_{\text{present}}}\right|$, 
between the present and literature values of drag coefficient $\delta_{\text{r}}(C_{\text{D}})\sim 2.75\%$ and $\sim 0.75\%$ for Newtonian unconfined ($\beta=\infty$) and non-Newtonian confined ($\beta=4$) 
flows. 
Based on our previous experiences \citep{Bharti2006,Bharti2007a,Bharti2007b,Sivakumar2006,Patnana2009,Patnana2010,Tian2014,Pravesh2019ijhmt}, such a small deviation is 
prone in numerical studies due to inherent characteristics of numerical techniques and methodologies used in related literature studies. The numerical results,  therefore, presented hereafter can be considered to be accurate within $\pm 1-2\%$.
%
\subsection{Onset of flow separation and wake formation} 
\begin{figure}[h]
\centering
	\subfloat[$\text{X}^{\text{c}}(n,\beta)$ ]{\includegraphics[width=0.42\linewidth]{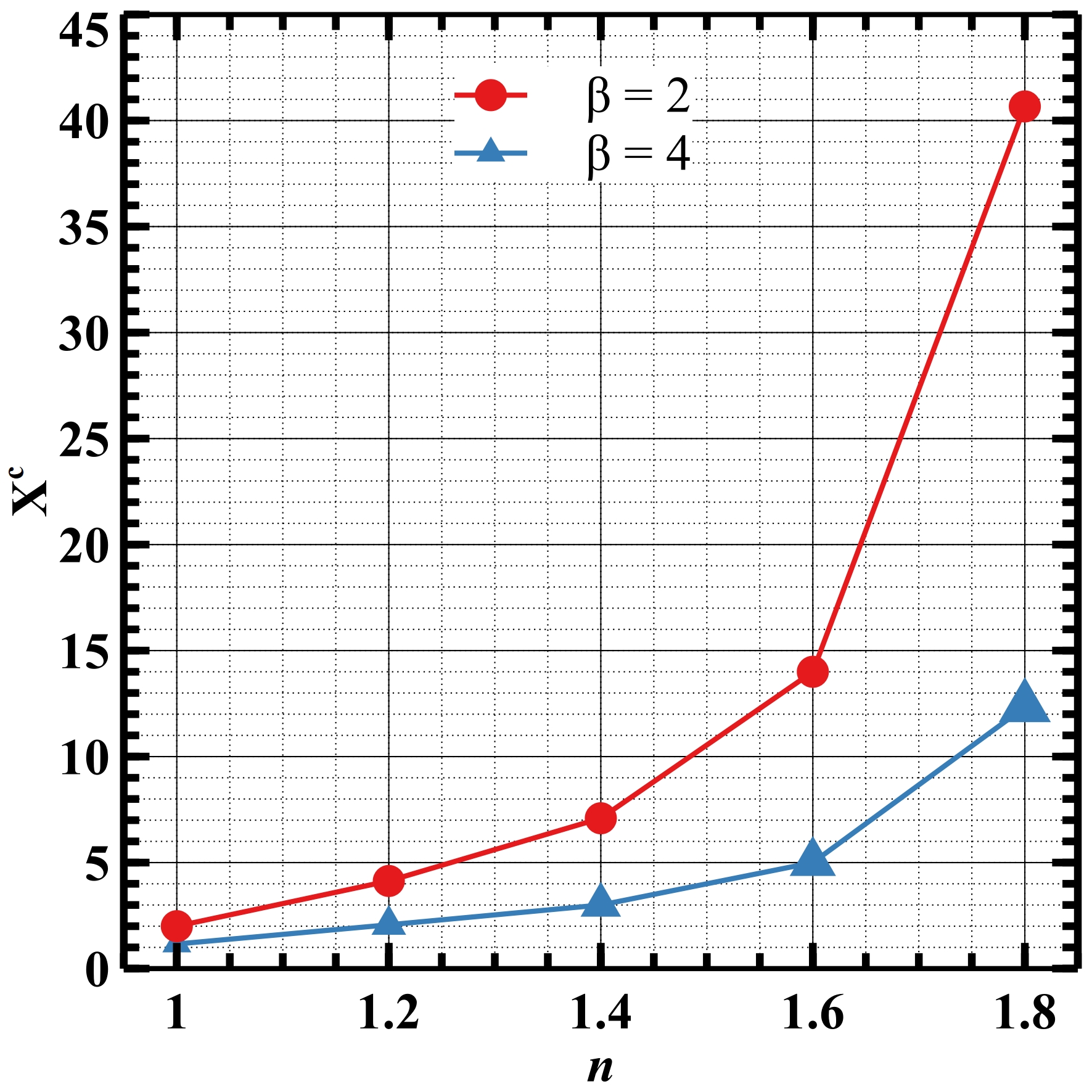}\label{fig:3a}}\qquad
\subfloat[$\text{Y}^{\text{c}}(n,\beta)$ ]{\includegraphics[width=0.42\linewidth]{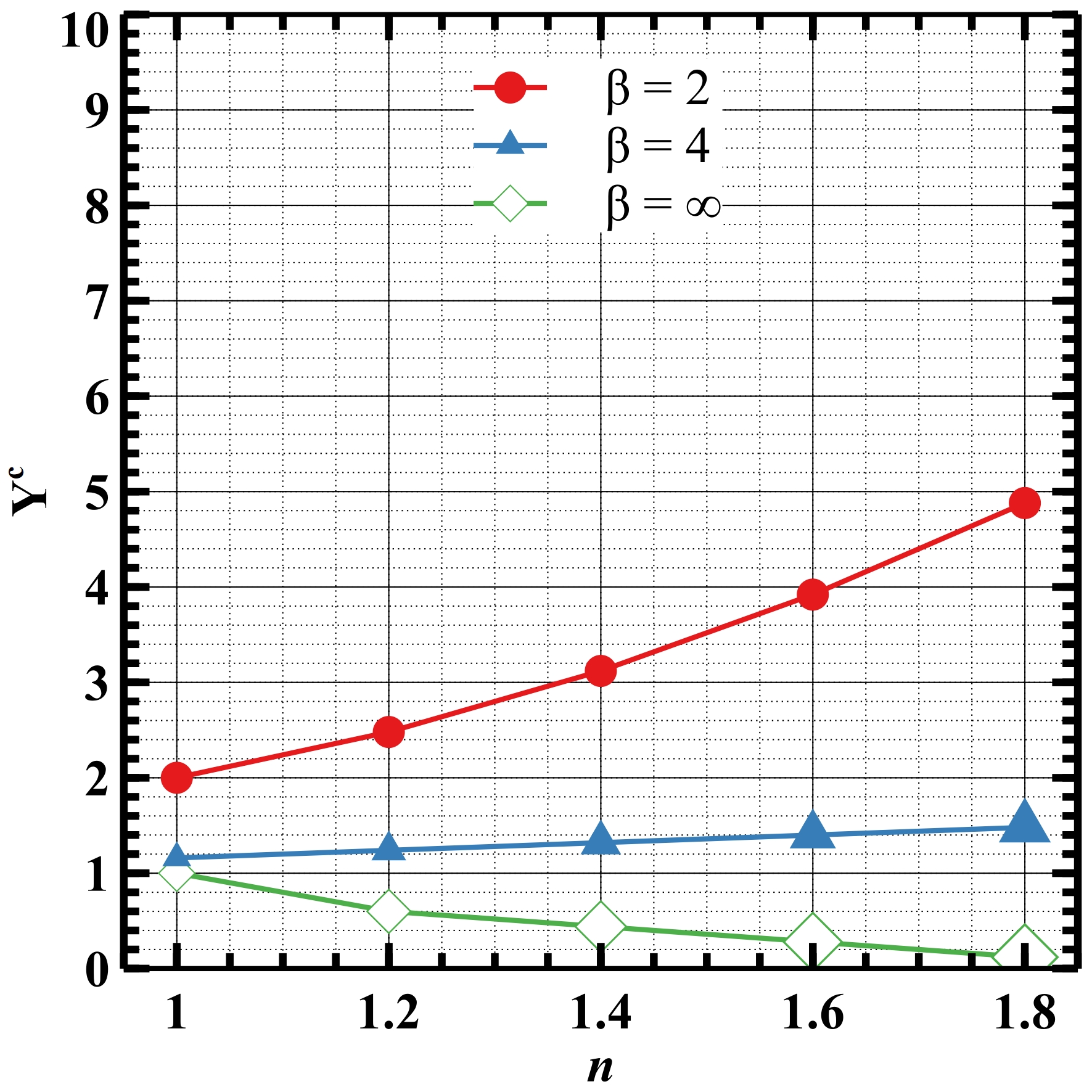}\label{fig:3b}}\\
\caption{Normalized critical Reynolds numbers ($\text{X}^{\text{c}}$ and $\text{Y}^{\text{c}}$) as a function of  power-law index ($n$) and wall blockage ($\beta$).}
\label{fig:3}
\end{figure}

\noindent 
This section presents the condition of transition from the creeping flow to two-dimensional (2-D) symmetric wake flow in terms of the lower critical Reynolds number ($Re^{\text{c}}$). The flow characteristics about both the horizontal ($x$, $y_{\text{c}}$) and vertical ($x_{\text{c}}$, $y$) axis passing through the center ($x_{\text{c}}$, $y_{\text{c}}$) of the cylinder are analyzed to locate the transitional conditions. The flow patterns in the creeping flow are known to be symmetric about both horizontal and vertical axis. Besides, both the pressure and viscous stress profiles over the surface of the cylinder \rev{emerge} to be symmetric.  
\begin{figure}[!h]
\centering
\begin{minipage}[b]{0.48\textwidth}
%
%
%
\subfloat[$Re_{l}=7.5$, $n=1.2$]{\includegraphics[width=0.4\linewidth]{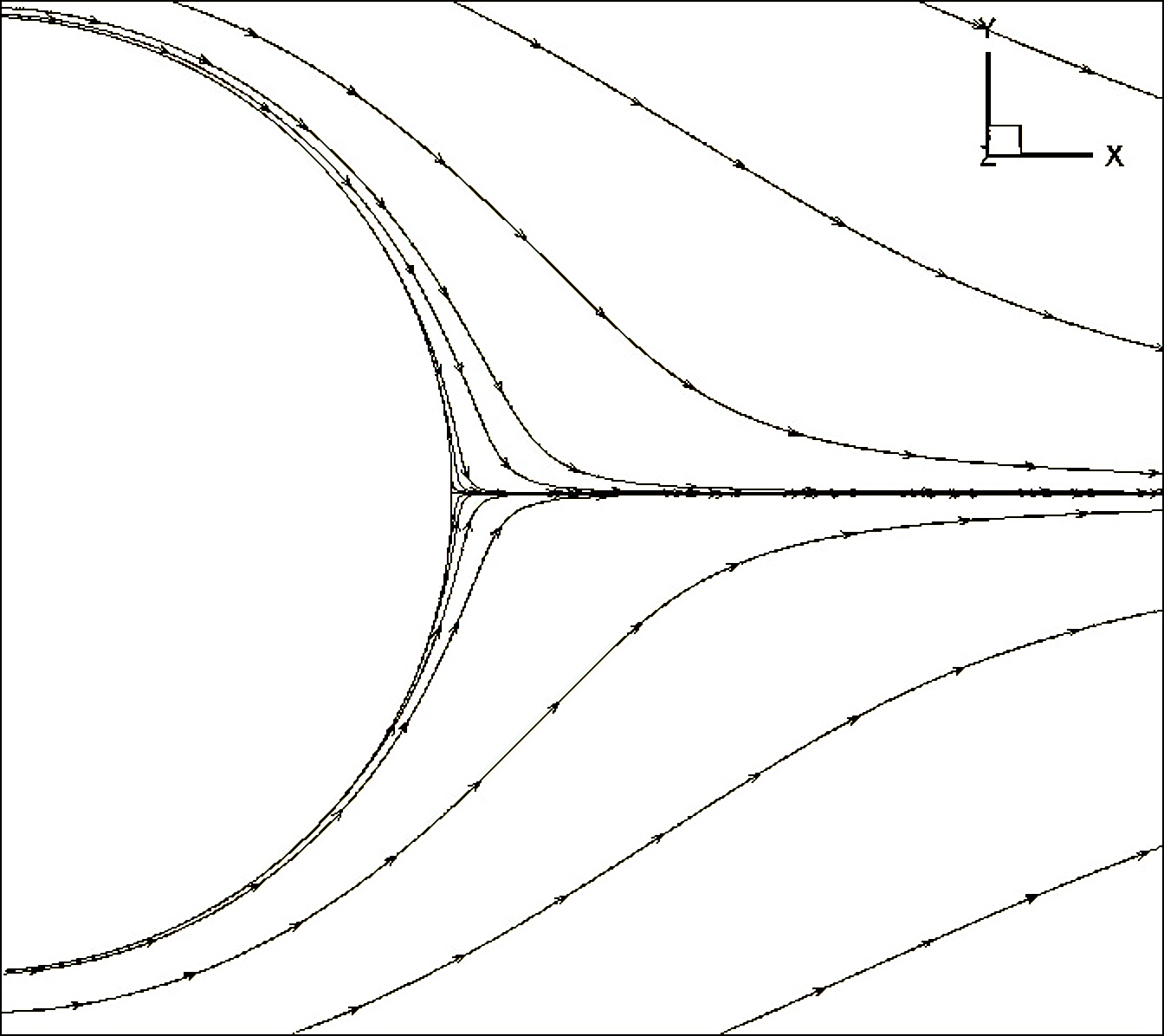}\label{fig:2b1}}\qquad
\subfloat[$Re_{u}=8$, $n=1.2$]{\includegraphics[width=0.4\linewidth]{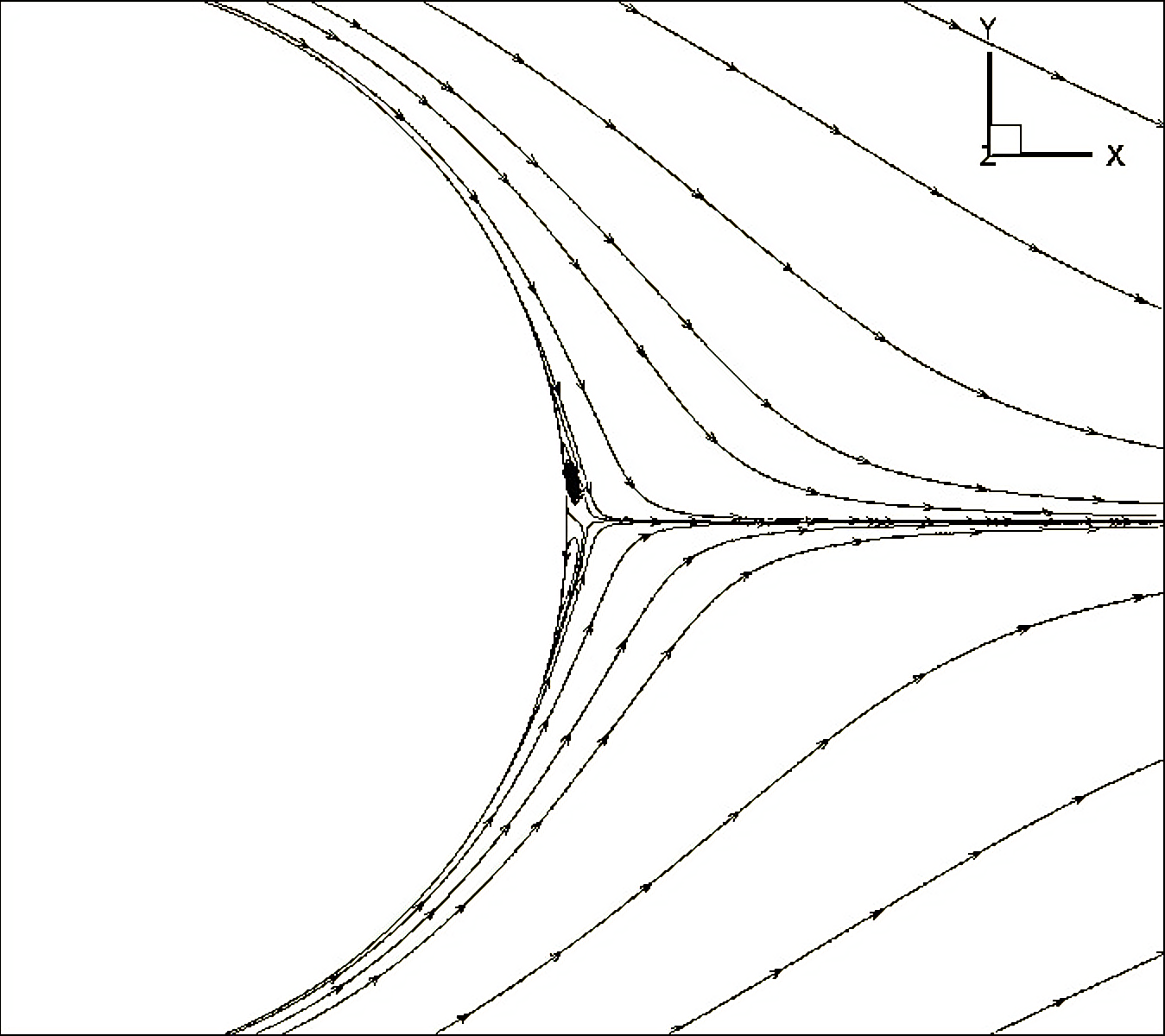}\label{fig:2b2}}\\
\subfloat[$Re_{l}=8$, $n=1.4$]{\includegraphics[width=0.4\linewidth]{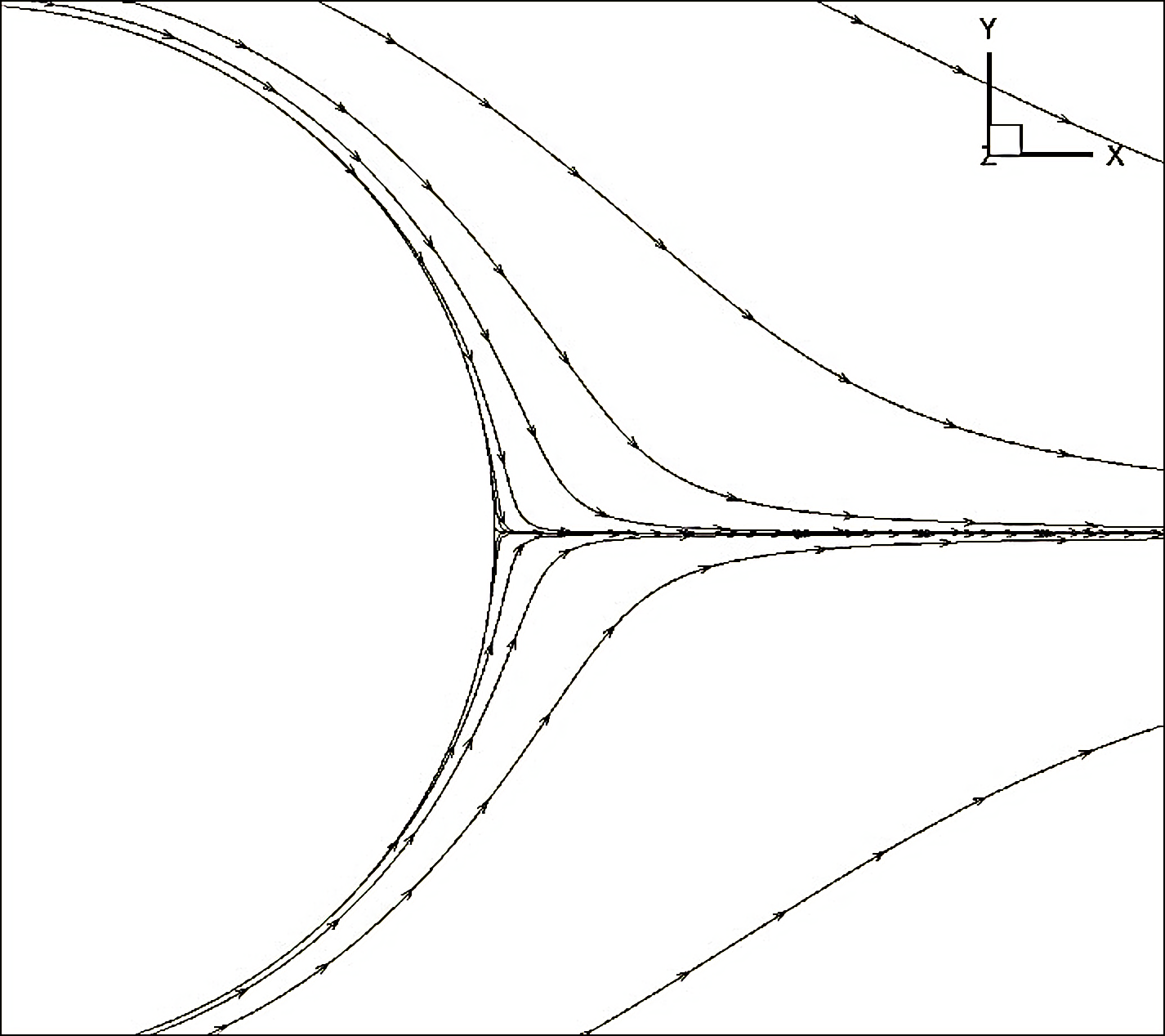}\label{fig:2c1}}\qquad
\subfloat[$Re_{u}=8.5$, $n=1.4$]{\includegraphics[width=0.4\linewidth]{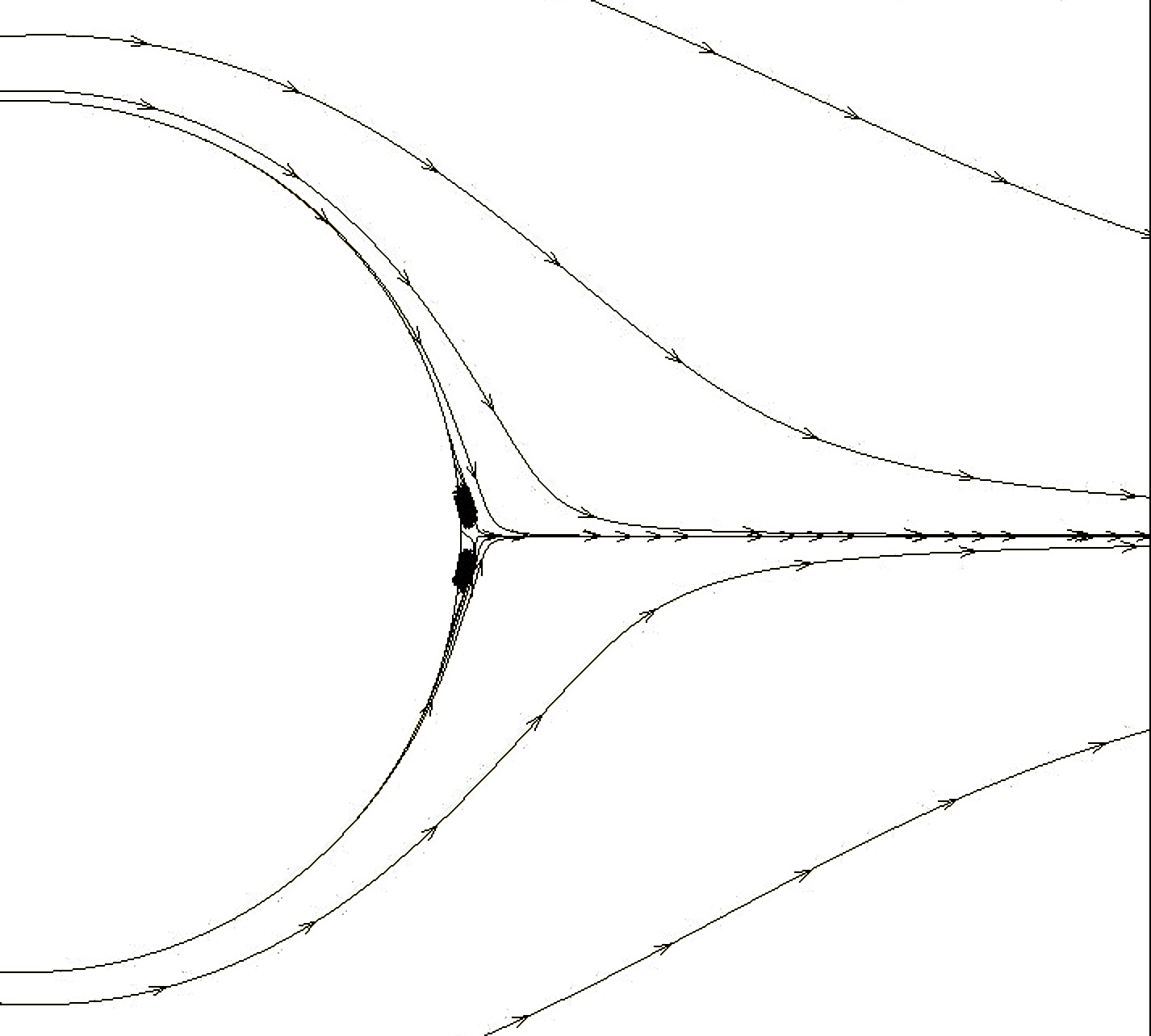}\label{fig:2c2}}\\
%
%
%
\subfloat[$Re_{l}=9$, $n=1.8$]{\includegraphics[width=0.4\linewidth]{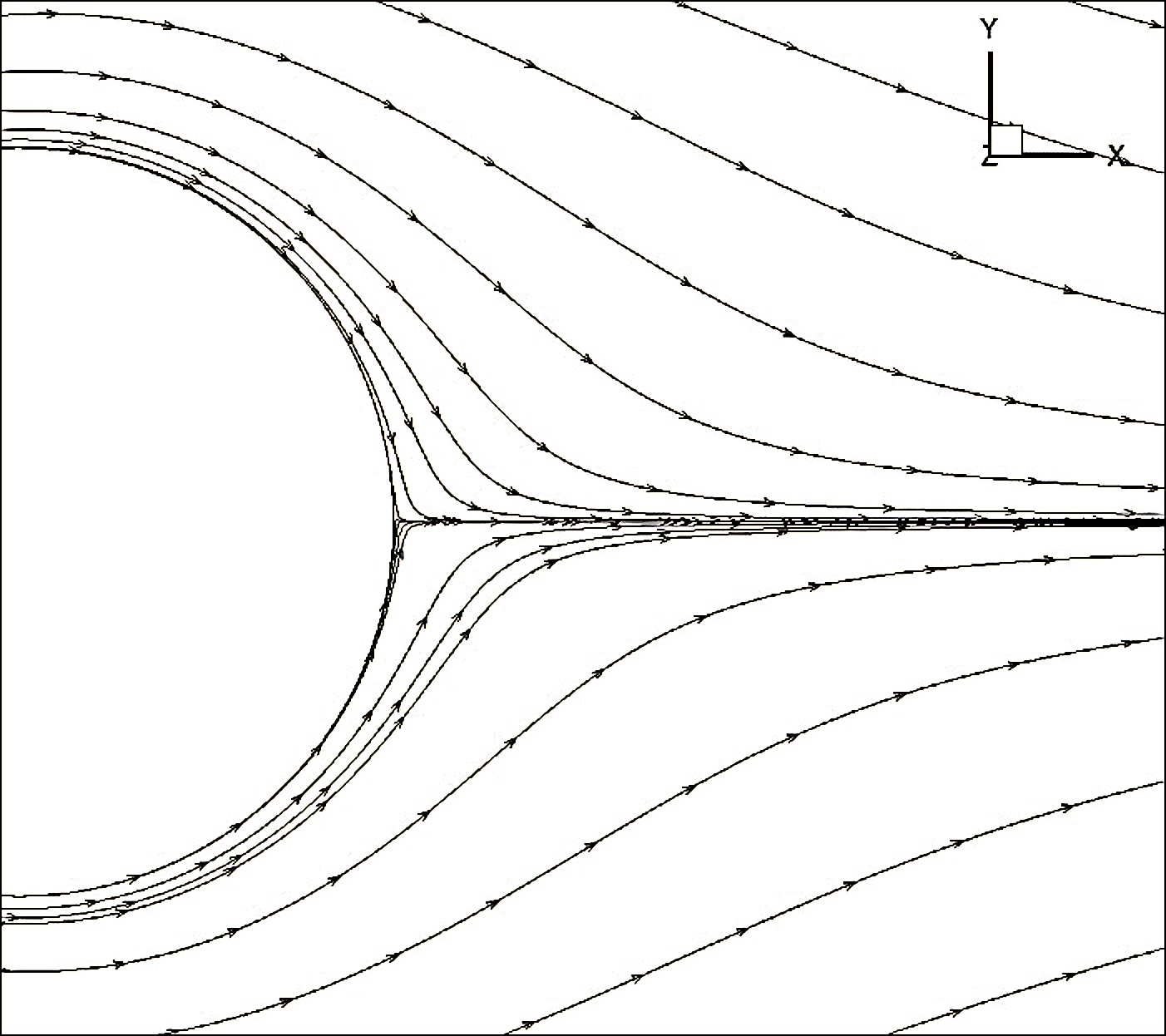}\label{fig:2e1}}\qquad
\subfloat[$Re_{u}=9.5$, $n=1.8$]{\includegraphics[width=0.4\linewidth]{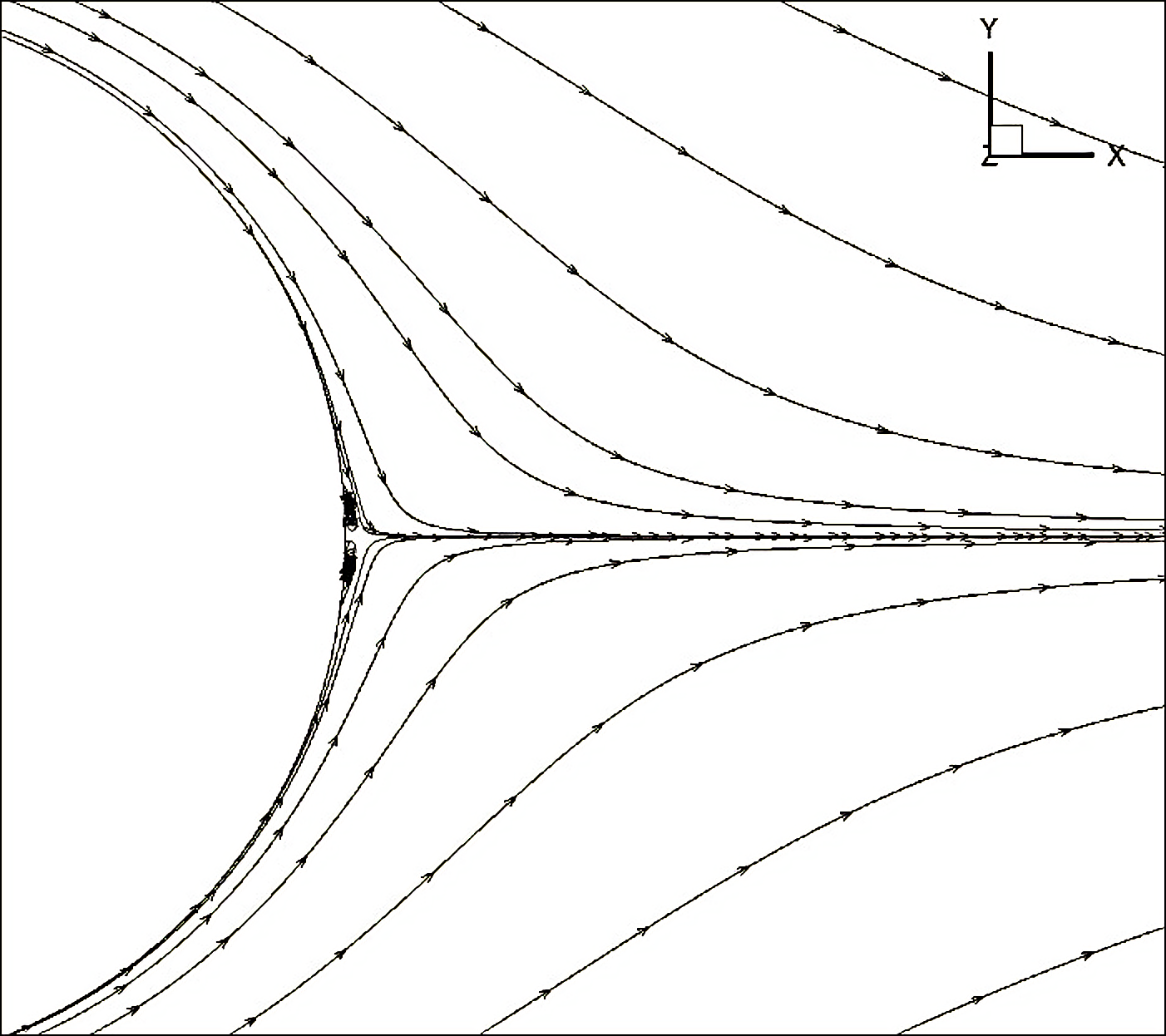}\label{fig:2e2}}\\
\centering (a - f) $\beta=4$
\end{minipage}
\begin{minipage}[b]{0.48\textwidth}
\subfloat[$Re_{l}=15$, $n=1.2$]{\includegraphics[width=0.4\linewidth]{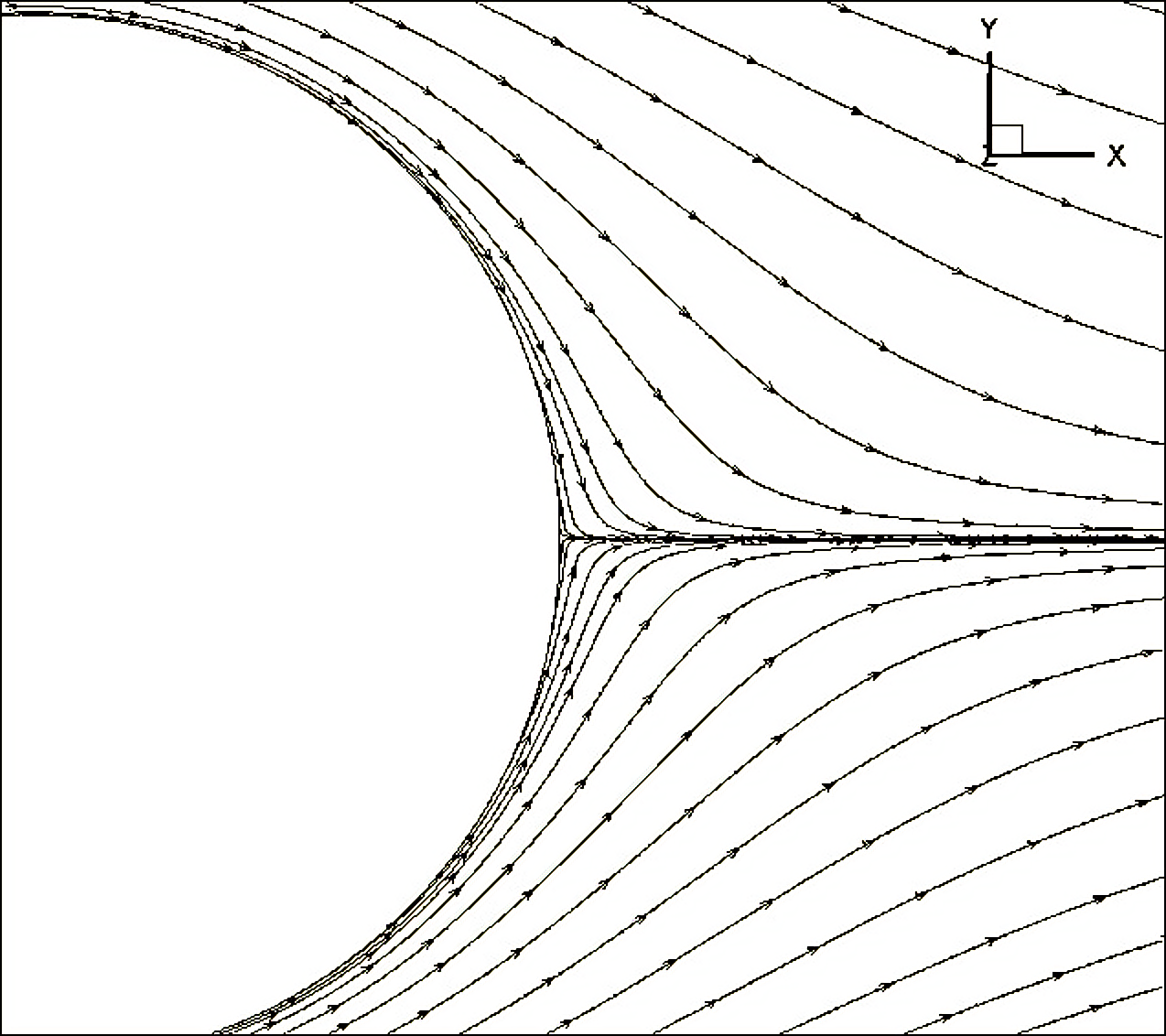}\label{fig:3b1}}\qquad
\subfloat[$Re_{u}=16$, $n=1.2$]{\includegraphics[width=0.4\linewidth]{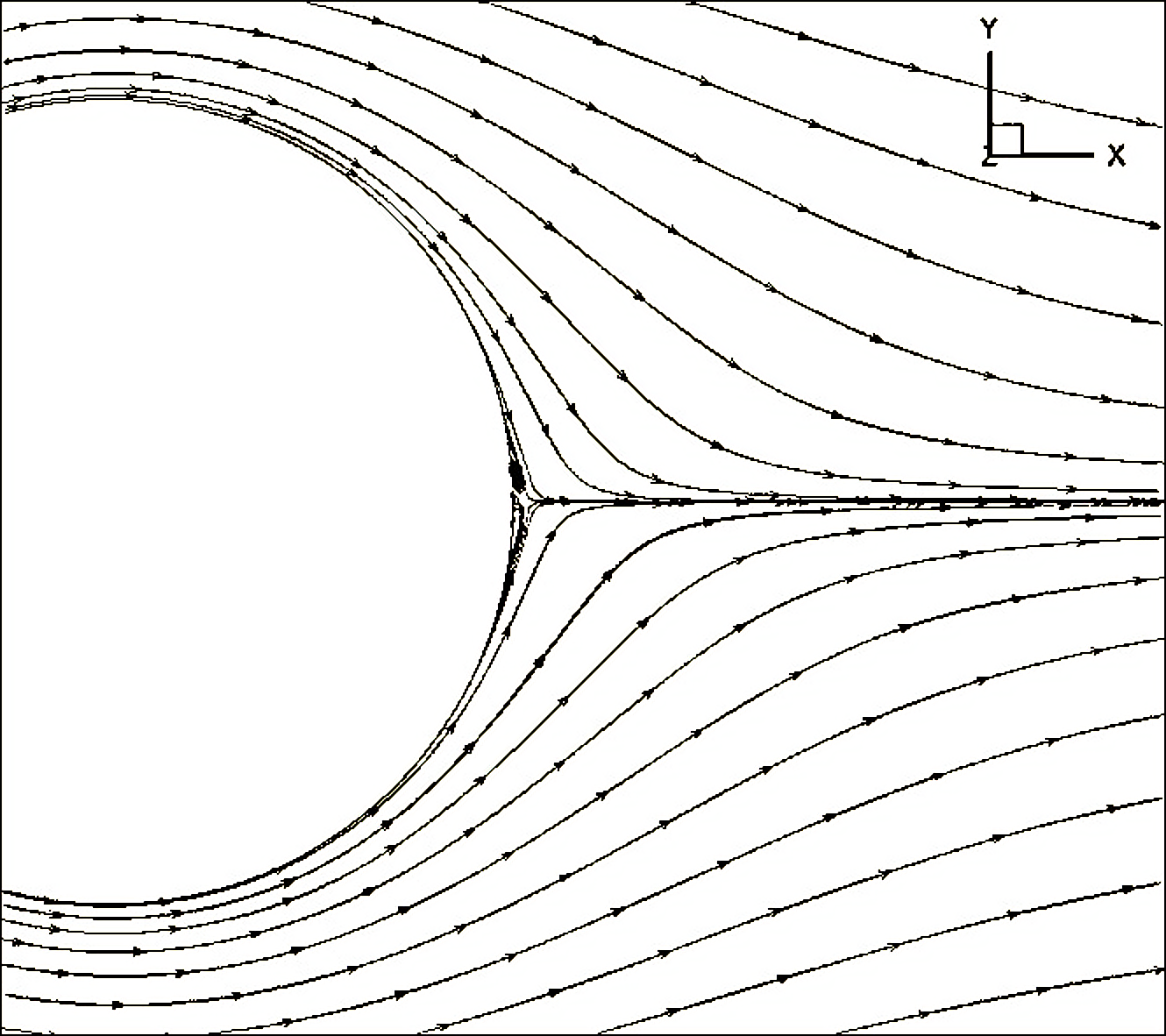}\label{fig:3b2}}\\
\subfloat[$Re_{l}=19$, $n=1.4$]{\includegraphics[width=0.4\linewidth]{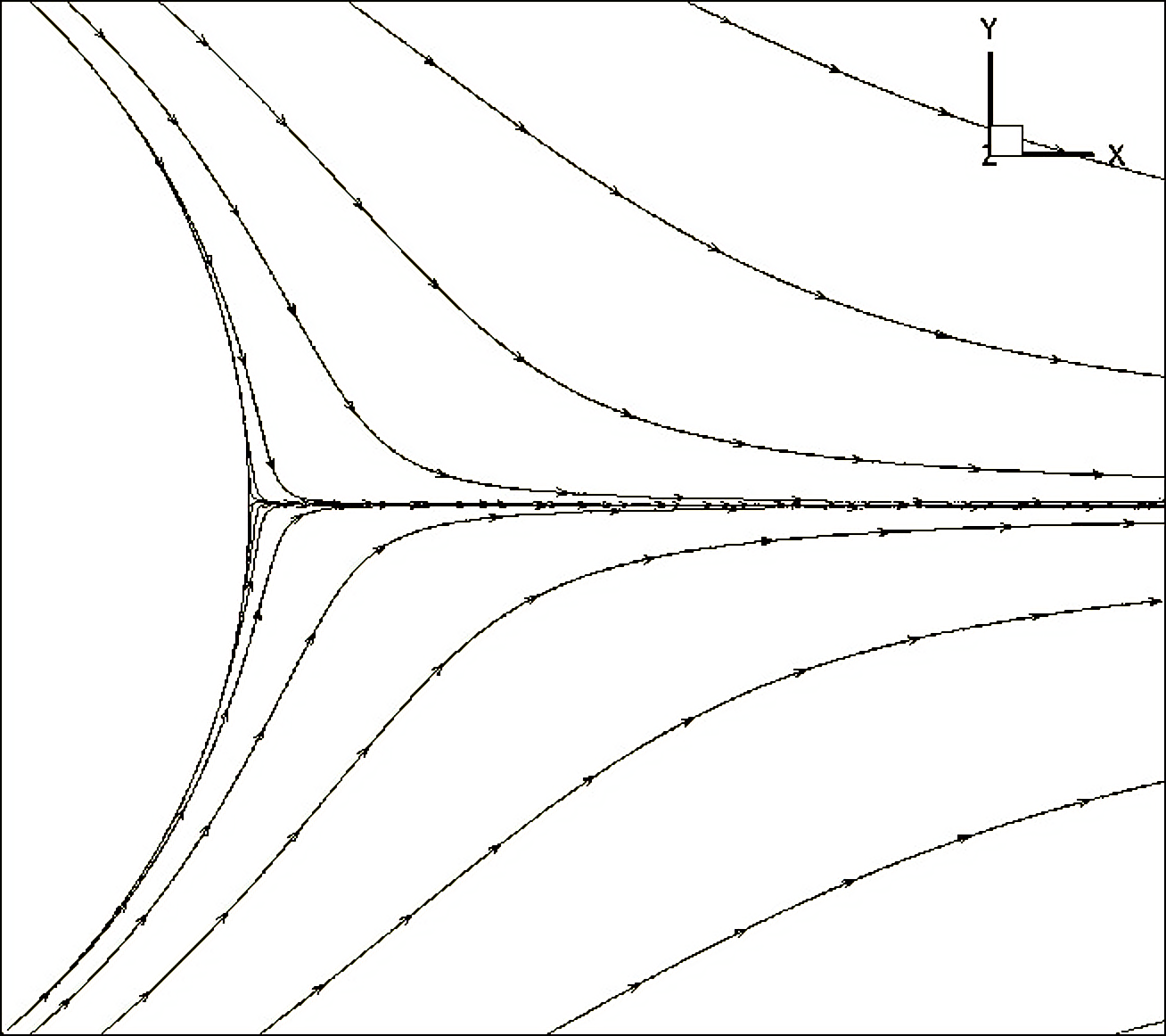}\label{fig:3c1}}\qquad
\subfloat[$Re_{u}=20$, $n=1.4$]{\includegraphics[width=0.4\linewidth]{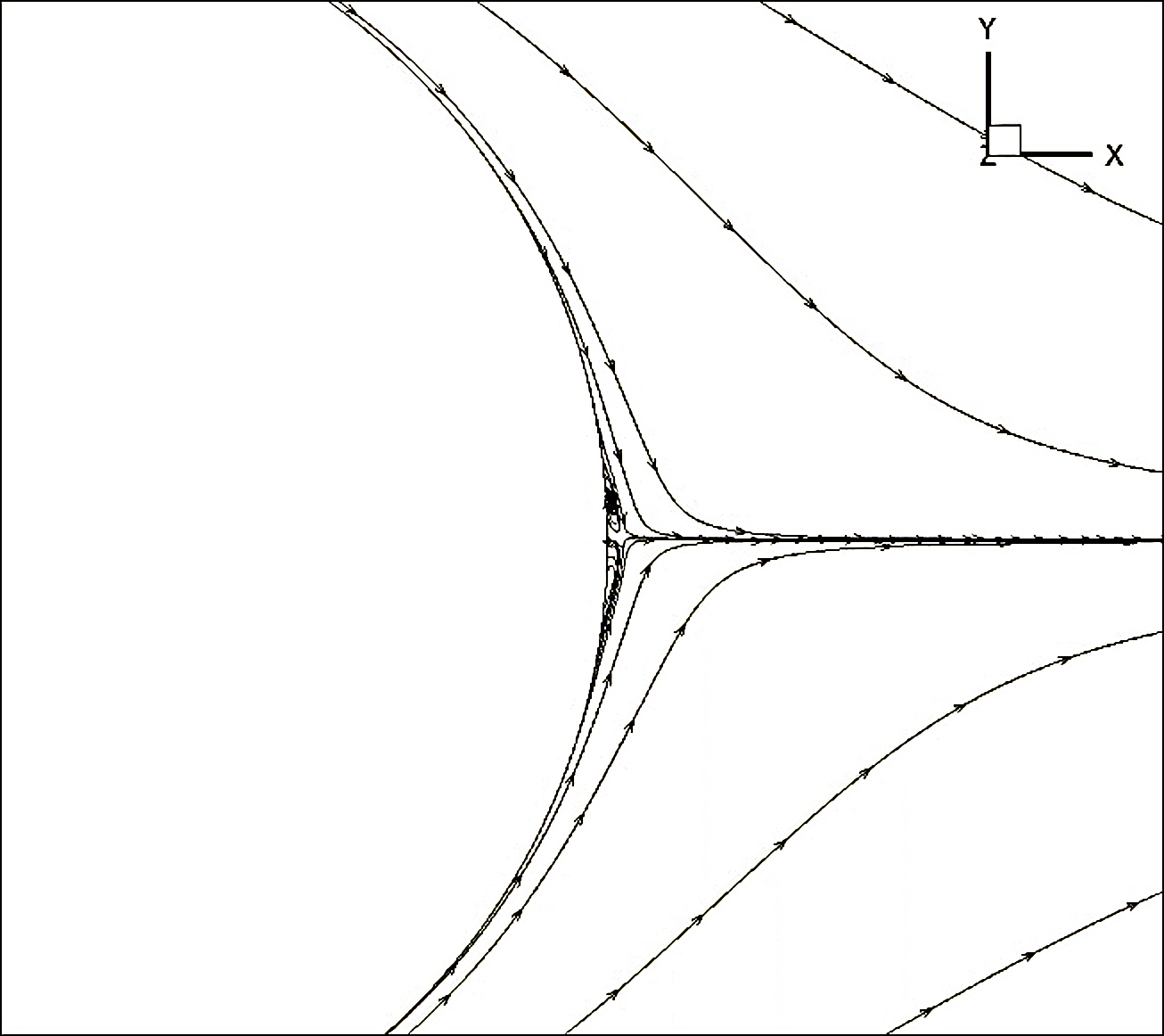}\label{fig:3c2}}\\
%
%
%
\subfloat[$Re_{l}=30$, $n=1.8$]{\includegraphics[width=0.4\linewidth]{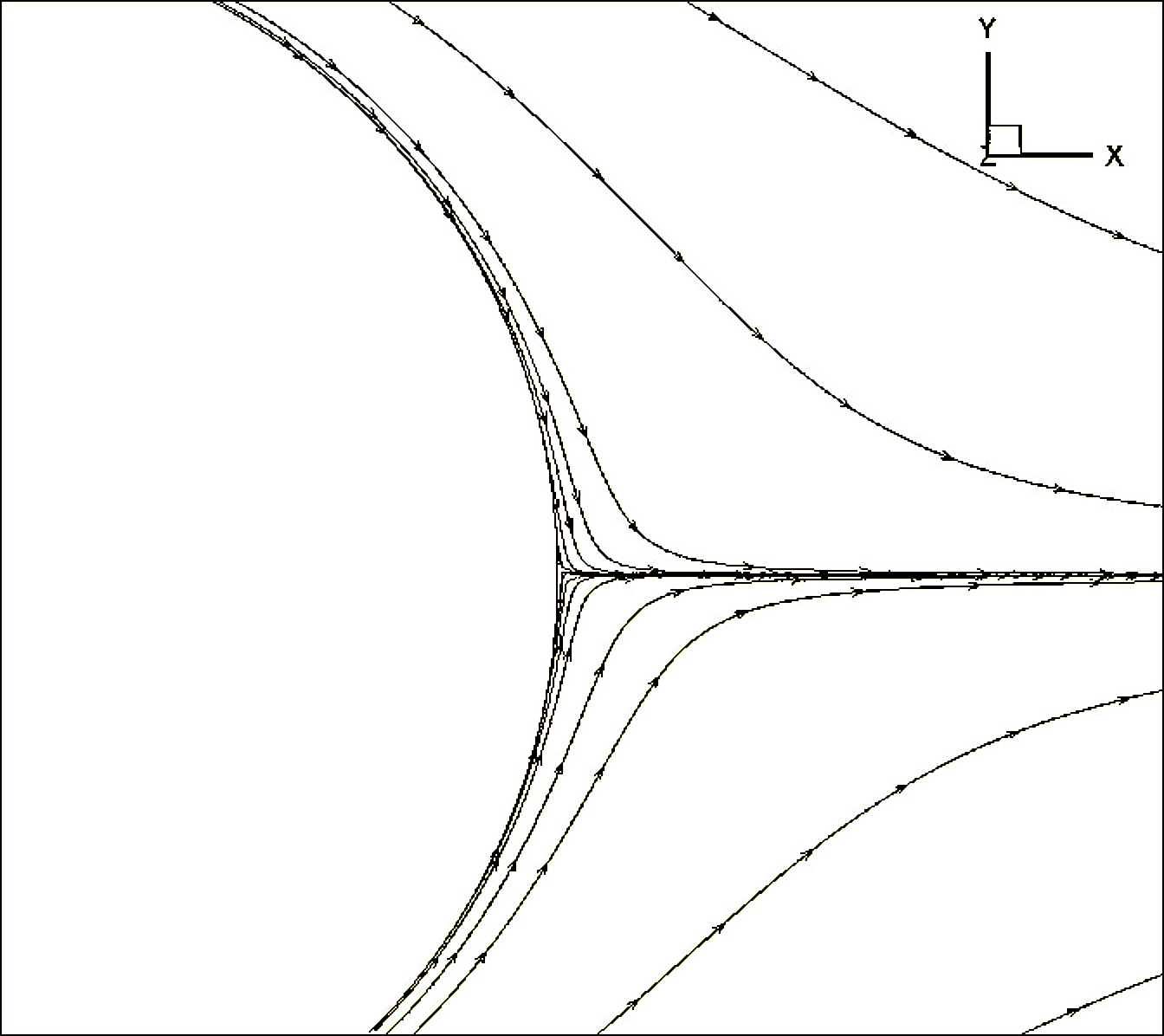}\label{fig:3e1}}\qquad
\subfloat[$Re_{u}=31$, $n=1.8$]{\includegraphics[width=0.4\linewidth]{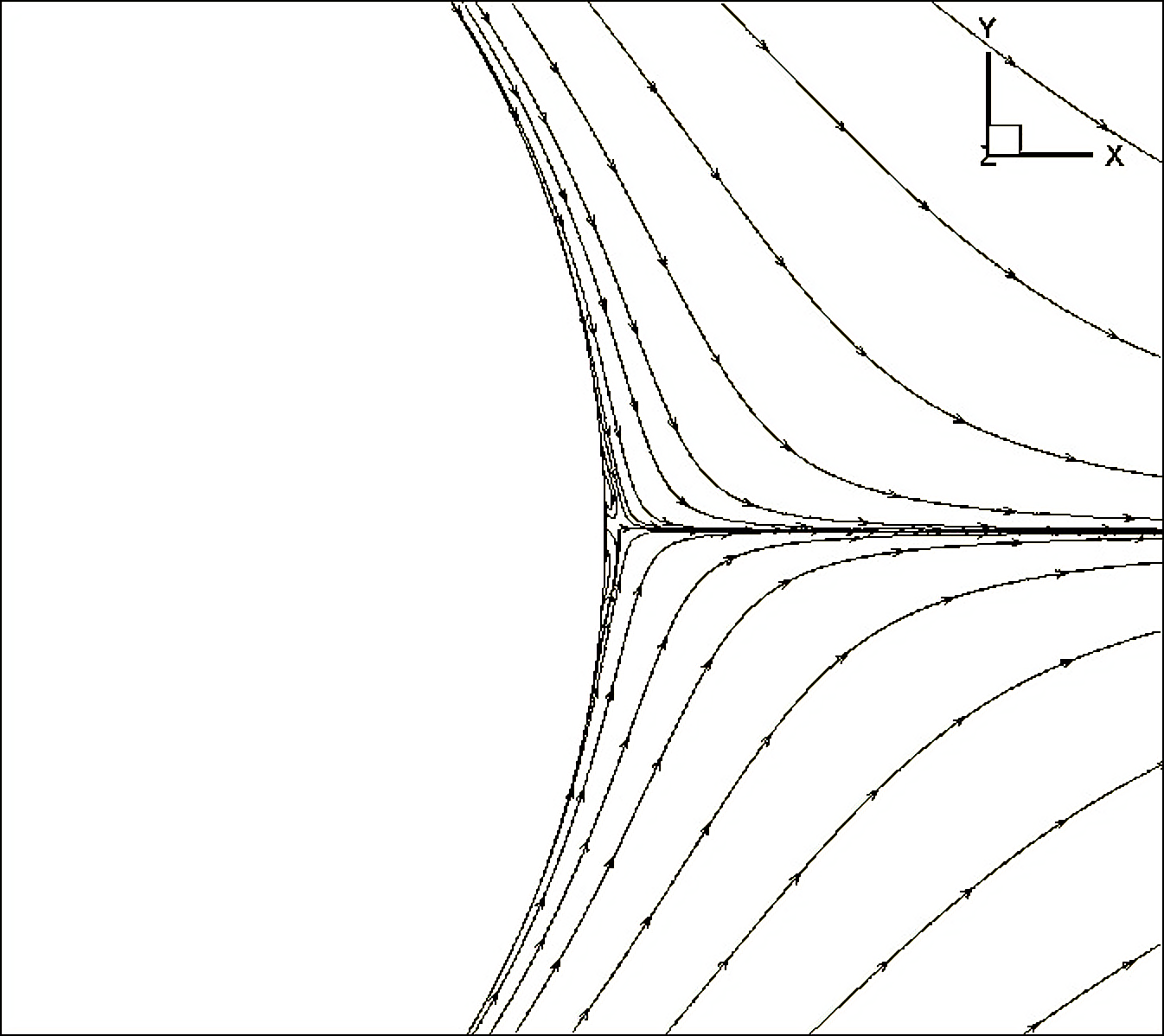}\label{fig:3e2}}\\
\centering (g - l) $\beta=2$
\end{minipage}
%
\caption{Streamline profiles representing the lower critical Reynolds numbers ($Re_{l}\le Re^c\le Re_{u}$) for various values of power-law index ($n$) for wall blockage of $\beta=4$  and 2.
The Reynolds numbers $Re_{l}$ and $Re_{u}$ indicate `no flow separation' and `flow separation', respectively.}
\label{fig:2}
\end{figure} 
As the flow transits from creeping to symmetric wake flow, streamline patterns, and pressure and viscous stresses over the surface of the cylinder show asymmetry about the vertical axis (i.e., in the fore and aft) of the cylinder.  In contrast, all the flow characteristics are symmetric about the horizontal axis, similar to the creeping flow. 
\rev{Furthermore, the pressure coefficient over the surface of cylinder remains positive ($C_{\text{p}}>0$) in creeping flow whereas it becomes zero ($C_{\text{p}}=0$) at the point of flow separation \citep{Bharti2006,Bharti2007a}. The friction coefficient (i.e., dimensionless wall shear stress) also equals to zero ($C_{\text{f}}=0$) at the point of separation.} 
\\\noindent
In this work, the dimensionless stream function ($\psi$) values adjacent to the cylinder, and pressure and friction coefficients ($C_{\text{p}}$ and $C_{\text{f}}$) over the surface of a cylinder are compared about the vertical axis (i.e., in the fore and aft) of the cylinder to identify the lower critical Reynolds number ($Re^{\text{c}}$).
The stream function value at the surface of a solid cylinder is assumed to be zero, $\psi=0$. The flow is believed to be without separation, i.e., creeping flow, for $\psi \le 10^{-5}$ adjacent to a cylinder \rev{and the pressure coefficient remains positive ($C_{\text{p}} \ge 0 $)  in the rear-side of the cylinder}. For larger values of $\psi > 10^{-5}$, onset of flow separation and symmetric wake formation is considered due to the loss of vertical (i.e., in the fore and aft) symmetry.
The value of Reynolds number ($Re$) at which stream function value changes from $\psi \le 10^{-5}$ to $\psi > 10^{-5}$ and the pressure profile transits from positive ($C_{\text{p}} > 0 $) to zero ($C_{\text{p}} < 10^{-5}$) is recorded as the lower critical Reynolds number  ($Re^{\text{c}}$), under otherwise identical conditions.
%
\fig\ref{fig:2} shows the streamline profiles schematically representing for the `no separation' (at $Re_{\text{l}}$) and `separation' (at $Re_{\text{u}}$) of the flow in the close vicinity behind the cylinder for a range of power-law index ($1\le n\le 1.8$) and wall blockage ($\beta=2$ and 4). The critical Reynolds number ($Re^{\text{c}}$) can, thus, be marked as the lowest point or appearance for the two-dimensional symmetric wake flow regime.
\\\noindent 
\rev{Based on the above discussed analysis,} the effect\rev{s} of power-law index ($n$) and wall blockage ($\beta$) on the  onset of \rev{flow separation and} wake formation in terms of critical Reynolds number ($Re_{\text{l}}\le Re^{\text{c}}\le Re_{\text{u}}$) have been recorded and presented in \tab\ref{Tab:4}. For the comparison purpose, the results for an unconfined ($\beta=\infty$) flow over a circular cylinder are also obtained and included in \tab\ref{Tab:4}, which are replicated and consistent with those reported in the literature \citep{Sivakumar2006}. 
%
%
\begin{table}
\centering
\caption{Critical Reynolds numbers ($Re_{\text{l}}\le Re^{\text{c}} \le Re_{\text{u}}$ and $Re_{\text{l}}\le Re_{\text{c}} \le Re_{\text{u}}$) as a function of power-law index ($n$) and wall blockage ($\beta$).}\label{Tab:4}
\fontsize{9}{13pt}\selectfont
\begin{tabular}{ccccccc}
\hline
& 	\multicolumn{3}{l|}{Lower critical Reynolds number ($Re^{\text{c}}$)}& 	\multicolumn{3}{l}{Upper critical Reynolds number ($Re_{\text{c}}$)} \\\cline{2-7}	
$n$	&$\beta=2$	&$\beta=4$	&$\beta=\infty$	&$\beta=2$	&$\beta=4$	&$\beta=\infty$	\\\hline
1	&\recl{12}{13}	&\recl{7.0}{7.5}	&\recl{6.0}{6.5}	&\recu{84}{85}	&\recu{70}{70.5}	&\recu{46}{47}	\\
1.2	&\recl{15}{16}	&\recl{7.5}{8.0}	&\recl{3.5}{4.0}	&\recu{149}{150}&\recu{88}{89}		&\recu{43}{44}	\\
1.4	&\recl{19}{20}	&\recl{8.0}{8.5}	&\recl{2.5}{3.0}	&\recu{219}{220}&\recu{106}{107}	&\recu{40}{41}	\\
1.6	&\recl{24}{25}	&\recl{8.5}{9.0}	&\recl{1.5}{2.0}	&\recu{345}{346}&\recu{156}{157}	&\recu{36}{37}	\\
1.8	&\recl{30}{31}	&\recl{9.0}{9.5}	&\recl{0.5}{1.0}	&\recu{449}{450}&\recu{179}{180}	&\recu{33}{34}	\\
\hline
\end{tabular} 
\end{table}
\tab\ref{Tab:4} shows that the critical Reynolds number ($Re^{\text{c}}$) increases, i.e., flow separation delays, with an increasing value of the power-law index ($n$) for a fixed wall blockage ($\beta$). Similarly, the flow separation is seen to delay with an increasing wall confinement (i.e., decreasing $\beta$) for a fixed value of the flow behaviour index ($n$).
The dependence of $Re^{\text{c}}$ on $n$ shown for the confined  ($\beta=2$ and 4) flows is, however, completely opposite to that for unconfined  ($\beta=\infty$) flow and $Re^{\text{c}}$ decreased with increasing $n$. 
\rev{The wall confinement is very likely stabilizing the local flow acceleration generated due to the cylinder and causes the delay in the flow separation for a given flow behaviour index ($n$)}, as shown elsewhere \citep{Bharti2007a,Bharti2007b} through the streamline and isotherm profiles.
It is noteworthy that there is `no creeping flow'\rev{, i.e., Stokes paradox,} for highly shear-thickening  ($n > 1.8$) fluids \citep{Tanner1993,MPaloka2001,Sivakumar2006} \rev{flow over an unconfined cylinder}, whereas the critical Reynolds number ($Re^{\text{c}}$) increases in confined flows 
with increasing value of the power-law index ($n$). For instance, the lower critical Reynolds number ($Re^{\text{c}}$) for unconfined ($\beta=\infty$) flow reduces from  $\sim{6.25}$ to $\sim{0.75}$ with increase in flow behaviour index ($n$) from 1 to 1.8. 
It thereby suggests that further increasing level of the shear-thickening ($n>1.8$) is expected to result in wake formation even \rev{at $Re\rightarrow 0$ and no appearance of the creeping flow (i.e., Stokes paradox)}. It is, however, not the case with confined flows (finite $\beta$) and $Re^{\text{c}}$ increased from $\sim{12.5}$ to $\sim{30.5}$ and  from $\sim{7.75}$ to $\sim{9.25}$ with increasing $n$ from 1 to 1.8 at $\beta=2$ and 4, respectively. 
\rev{It is attributed to the complex interplay between the inertial and frictional forces. The inertial force ($\propto\mathbf{u}^2$) remains constant whereas the viscous force ($\propto\mathbf{u}^{n}$) increases with increasing flow behaviour index ($n$) for a fixed blockage ratio ($\beta$). The viscous effects remain confined in the thin hydrodynamics boundary layer near the solid walls wherein both viscous and inertial forces are of the same order. 
The boundary layer thickness for the flow of a power-law fluid over the flat surface is also known  to increase with increasing $n$ and decreasing $Re$ \citep{Raju2015}. 
%
%
Furthermore, the minimum flow area between the channel wall and cylinder surface reduces with decreasing blockage ratio ($\beta$, i.e., increasing confinement) which in turn enhances the maximum local flow velocity ($\mathbf{u}_{\text{max}}$) and thereby enhancement of the local Reynolds number ($Re$). 
}
\\\noindent 
\rev{The functional dependence of the lower critical Reynolds $Re^c (n, \beta)$ is presented through the statistical analysis of the numerical data (shown in \tab\ref{Tab:4}) to broaden the usefulness in the design and engineering and expressed by \eqn(\ref{recnb}).} 
\begin{equation}
	Re^c(n, \beta)= a_4n^4 + a_3n^3+a_2n^2+a_1n+ (a_0 \pm \Delta)
	\qquad \text{for}\quad 2\le \beta \le \infty,\quad\text{and}\quad 1\le n\le 1.8	
\label{recnb}
\end{equation}
\rev{The coefficients ($a_0$ to $a_4$ and $\Delta$) appearing in the above predictive correlation (\eqn\ref{recnb}) are noted in \tab\ref{Tab:5}.} 
In comparison to an unconfined ($\beta=\infty$) flow wherein $Re^{\text{c}}$ have shown quartic (i.e., 4th order) dependence on $n$,  it shows linear and quadratic dependencies on $n$ for $\beta=4$ and 2, respectively (see \eqn\ref{recnb} and \tab\ref{Tab:5}).
\begin{table}
\centering
\caption{Predictive correlation coefficients.}\label{Tab:5}
\fontsize{10}{13pt}\selectfont
\begin{tabular}{ccrrrrrr}
\hline
&$\beta$ &$a_0$ & $a_1$&$a_2$ &$a_3$ &$a_4$& $\Delta $\\ \hline
$Re^{\text{c}}$(\eqn\ref{recnb})& $\infty$ &198.75  & -520.6250 & 523.4375& -234.3750 & 39.0625 & 0.25\\
& 4&4.75   &2.5000 & 0&0 &0 & 0.25\\
& 2&12.50   & -12.5000 & 12.5000& 0&0 & 0.50\\\hline
$Re_{\text{c}}$ (\eqn\ref{rec_nb})&$\infty$ & -360.50 & 770.4167&859.3750 &427.0833 &78.1250 & 0.50\\
& 4& -7541.00 & 23239.0000& -26280.0000& 13029.0000& -2376.3021 &0.50\\
& 2& - 10980.00 & 33384.0000& -37491.0000 &18531.0000 &-3359.3750 & 0.50\\
\hline
\end{tabular} 
\end{table}
\\\noindent 
\rev{Further, the relative impacts of flow behaviour index ($n$) and wall blockage ($\beta$) on the onset of wake formation are analysed by normalizing} the critical Reynolds number ($Re^{\text{c}}$) with respect to (a) an unconfined flow of non-Newtonian fluids ($\text{X}^{\text{c}}$), and (b) an unconfined flow of Newtonian fluids ($\text{Y}^{\text{c}}$), as defined by \eqn(\ref{nrec}). 
\begin{equation}
\text{X}^{\text{c}}=\frac{Re^{\text{c}}(n,\beta)}{Re^{\text{c}}(n,\infty)}\qquad\text{and}\qquad
\text{Y}^{\text{c}}=\frac{Re^{\text{c}}(n,\beta)}{Re^{\text{c}}(1,\infty)}
	\label{nrec} 
\end{equation}
\noindent\figs\ref{fig:3a} and \ref{fig:3b} depict the complex dependence of the normalized critical Reynolds number ($\text{X}^{\text{c}}$ and $\text{Y}^{\text{c}}$) on the dimensionless parameters ($n$ and $\beta$).
%
%
%
Qualitatively, the normalized factors ($\text{X}^{\text{c}}$ and $\text{Y}^{\text{c}}$) are seen to enhance, i.e., flow separation tends to delay, with increase in both the flow behaviour index ($n$) as well as the wall blockage ratio ($\beta$). The influences of dimensionless parameters ($n$ and $\beta$) are stronger for highly shear-thickening ($n\gg1$) fluids in comparison to those seen for Newtonian and mildly shear-thickening ($n\le 1.4$) fluids.  Similarly, the normalized factors ($\text{X}^{\text{c}}$ and $\text{Y}^{\text{c}}$) are greatly enhanced at small values of wall blockage ($\beta$). 
For instance, a factor $\text{X}^{\text{c}}=40.67$ and 12.33 is noted for $\beta=2$ and 4, respectively at $n=1.8$ against a factor $\text{X}^{\text{c}}=2$ and 1.16 at $n=1$. 
Qualitatively similar trends are also observed in case of the normalized factor $\text{Y}^{\text{c}}$.  For instance, a factor of $\text{Y}^{\text{c}}=4.88$ and 1.48 is noted for $\beta=2$ and 4, respectively at $n=1.8$ against a factor of $\text{Y}^{\text{c}}=2$ and 1.16 at $n=1$.
\\\noindent
The complex influences {of flow behvaiour index and wall blockage} on the wake formation, observed in this section, are also expected to influence the onset of wake instability. The subsequent section thus explores their effects on the onset of wake instability. 
\subsection{Onset of wake instability} 
\noindent 
This section presents the onset of wake instability, i.e., the condition of transition from the two-dimensional `symmetric'  to `asymmetric' wake flow regime in terms of the upper critical Reynolds number ($Re_{\text{c}}$). {In addition to the visualization of streamlines profile,} the value of lift coefficient ($C_{\text{L}}$) over a cylinder has been analyzed to demarcate the transition  from steady `symmetric' to `asymmetric' wake formation regime.  In particular, the wake formation is \rev{considered to be} the steady and symmetric for the value of lift coefficient approaching to zero ($C_L\le 10^{-4}$). 
%
%
\begin{figure}[!h]
\centering
\begin{minipage}[b]{0.48\textwidth}
	%
	%
	%
	\subfloat[$Re_{l}=88$, $n=1.2$]{\includegraphics[width=0.4\linewidth]{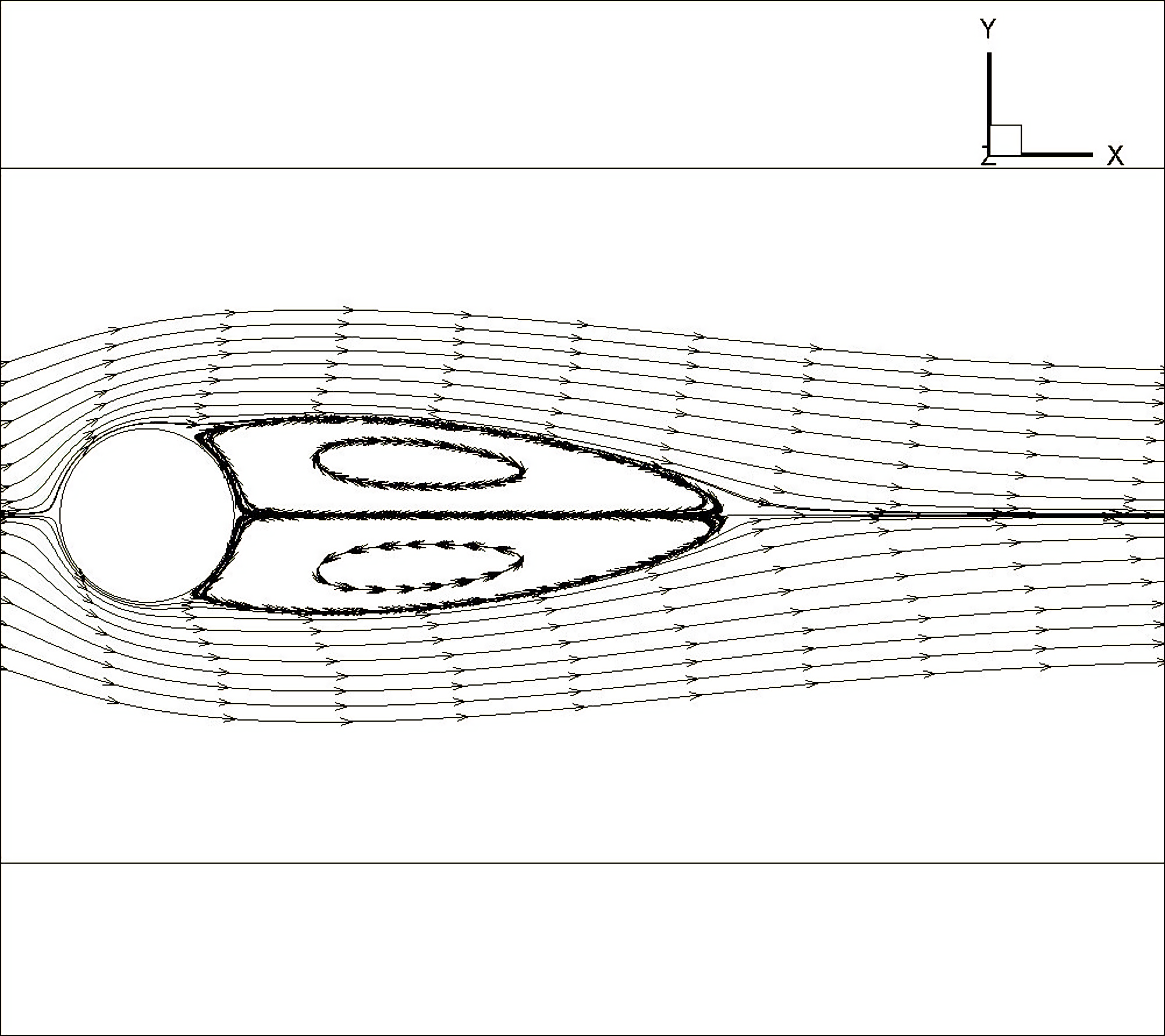}\label{fig:5b1}}\qquad
	\subfloat[$Re_{u}=89$, $n=1.2$]{\includegraphics[width=0.4\linewidth]{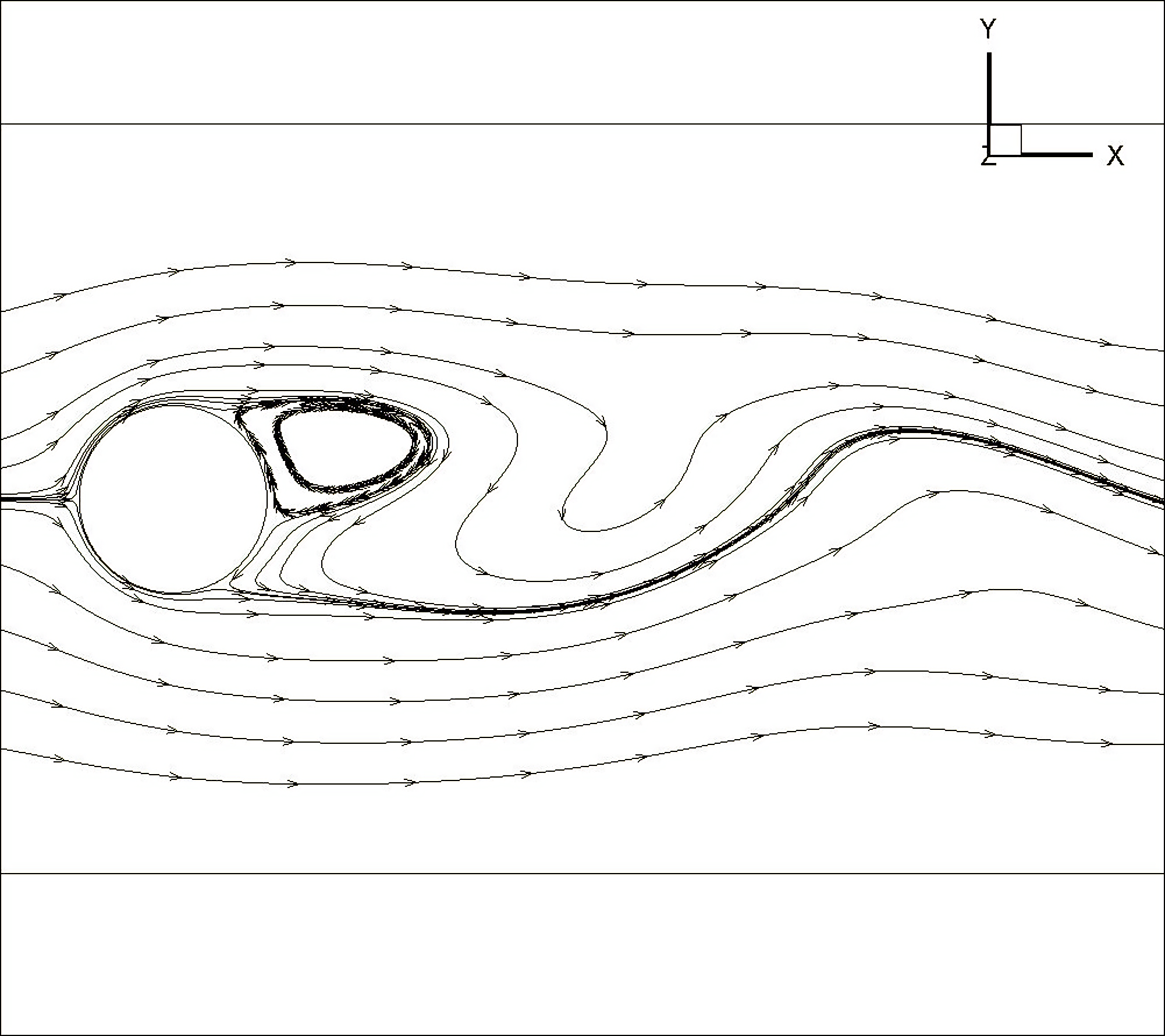}\label{fig:5b2}}\\
	\subfloat[$Re_{l}=106$, $n=1.4$]{\includegraphics[width=0.4\linewidth]{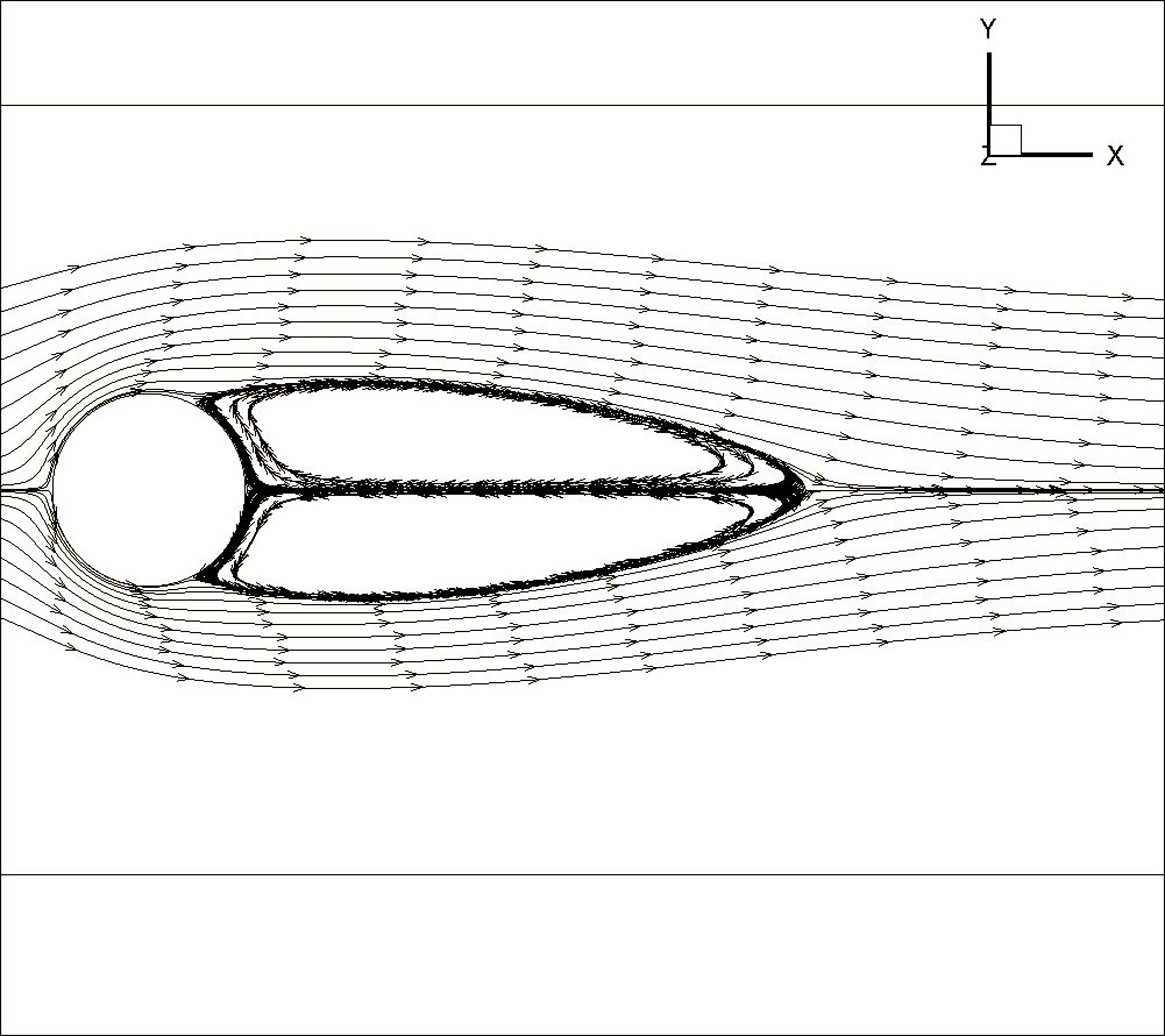}\label{fig:5c1}}\qquad
	\subfloat[$Re_{u}=107$, $n=1.4$]{\includegraphics[width=0.4\linewidth]{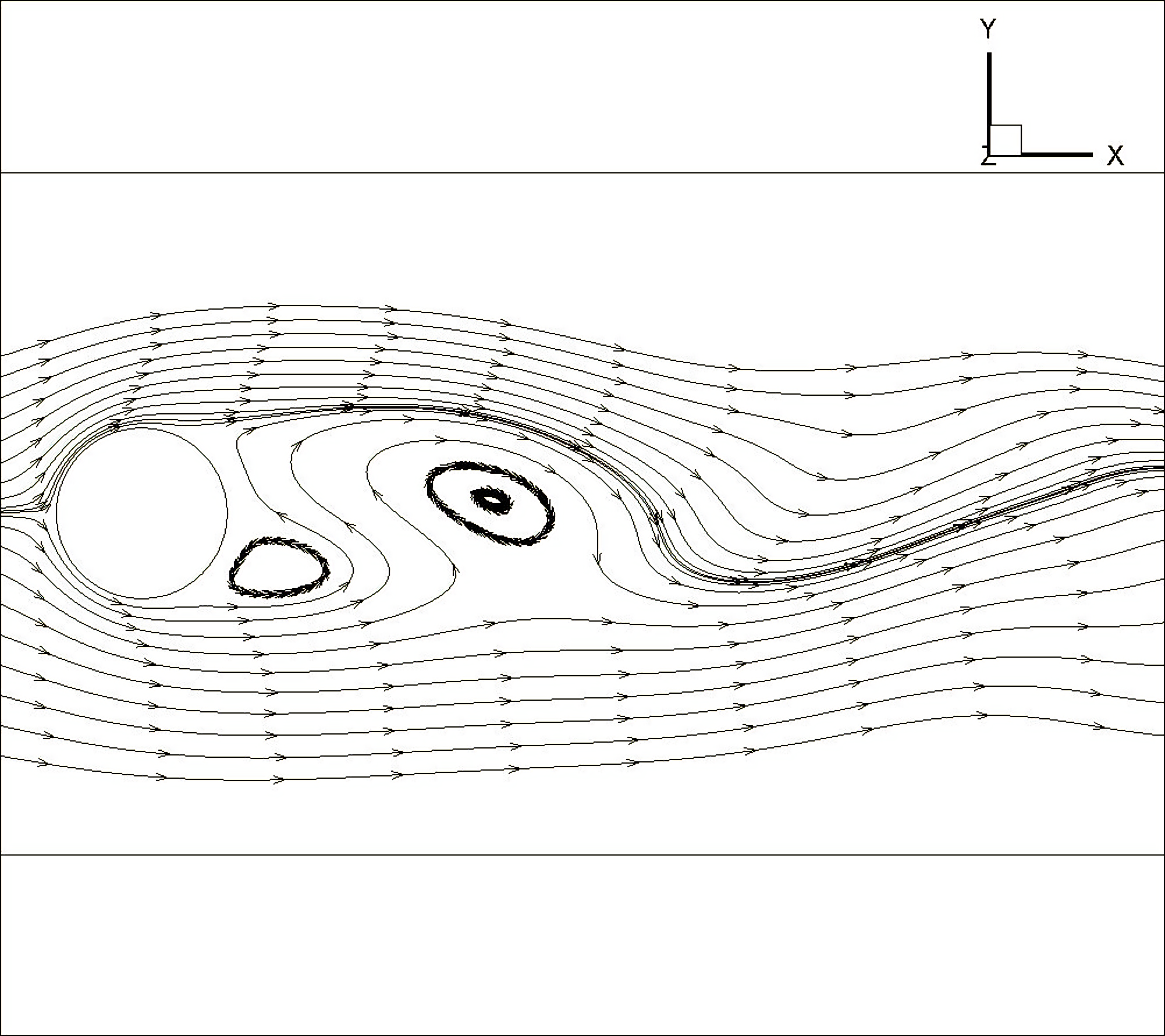}\label{fig:5c2}}\\
	%
	%
	%
	\subfloat[$Re_{l}=179$, $n=1.8$]{\includegraphics[width=0.4\linewidth]{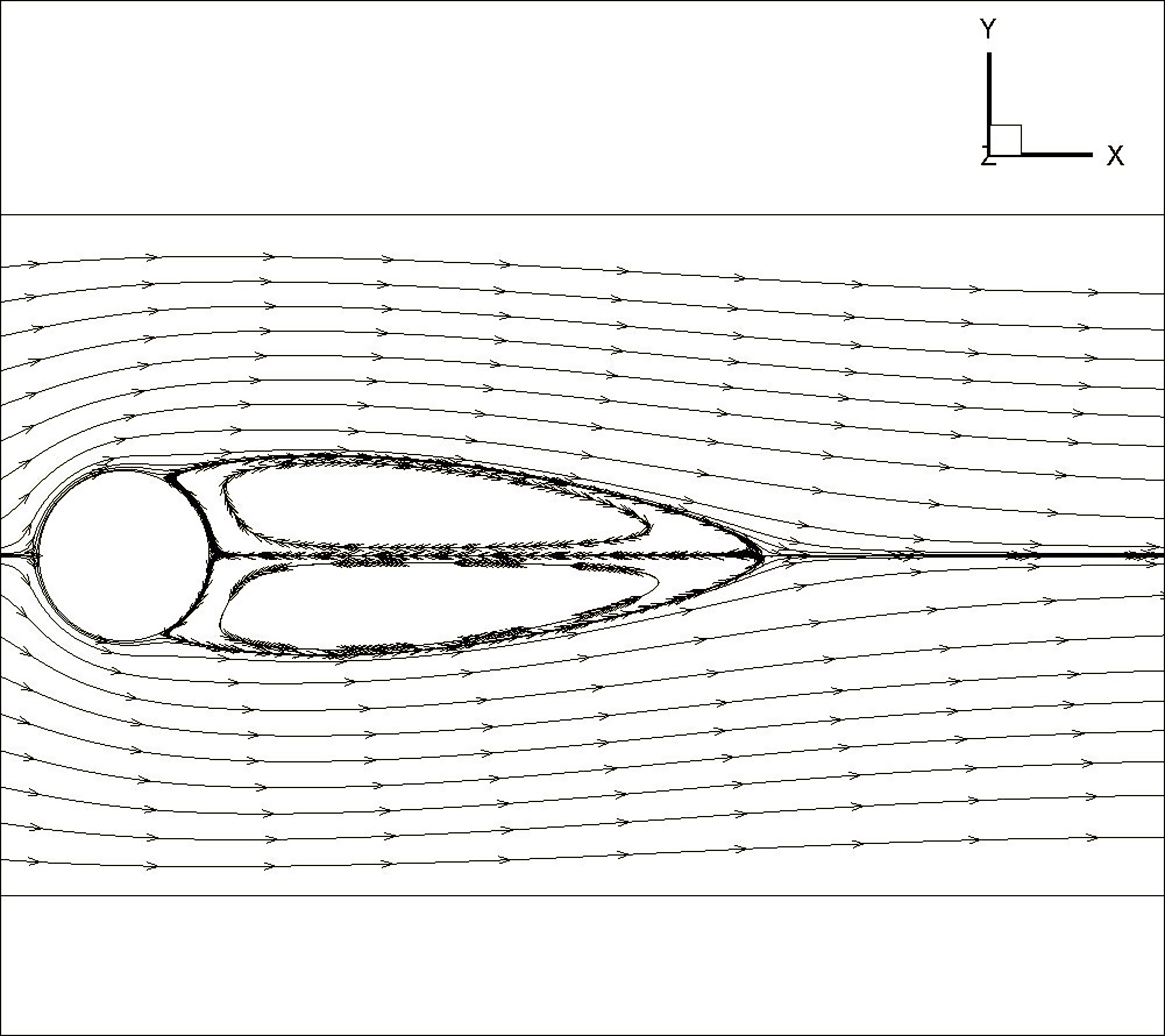}\label{fig:5e1}}\qquad
	\subfloat[$Re_{u}=180$, $n=1.8$]{\includegraphics[width=0.4\linewidth]{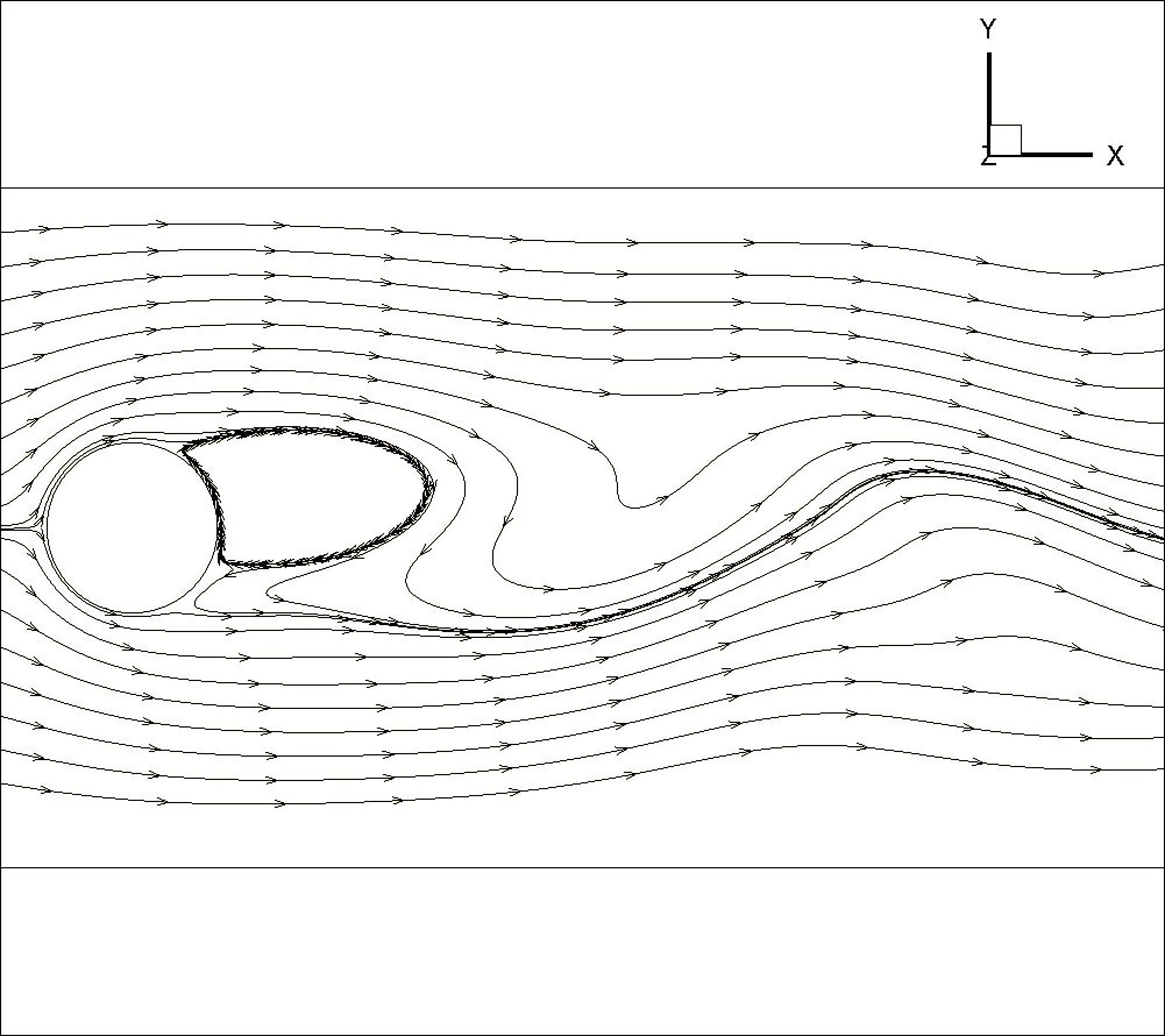}\label{fig:5e2}}\\
	\centering (a - f) $\beta=4$
\end{minipage}
\begin{minipage}[b]{0.48\textwidth}
	\subfloat[$Re_{l}=149$, $n=1.2$]{\includegraphics[width=0.4\linewidth]{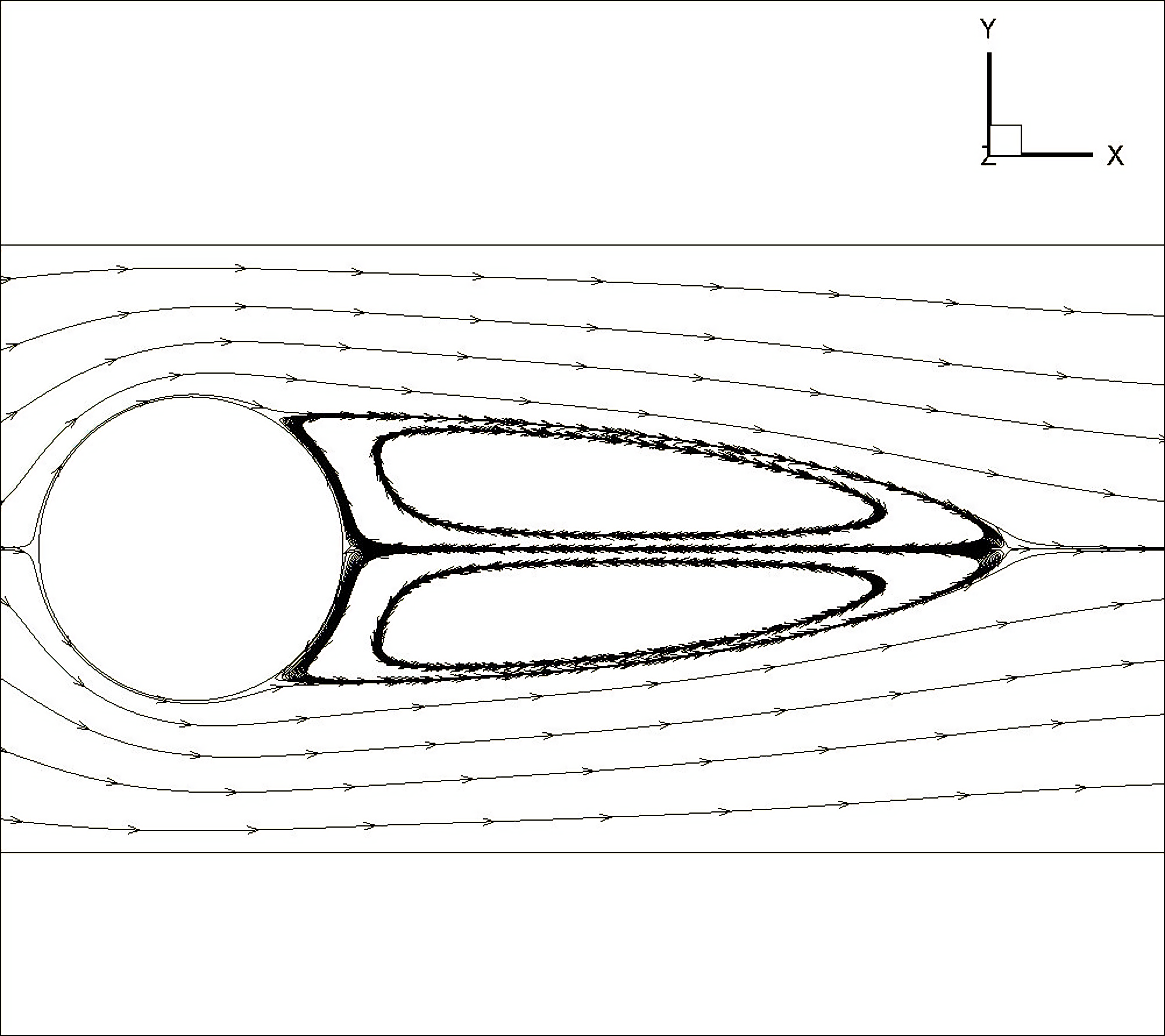}\label{fig:6b1}}\qquad
	\subfloat[$Re_{u}=150$, $n=1.2$]{\includegraphics[width=0.4\linewidth]{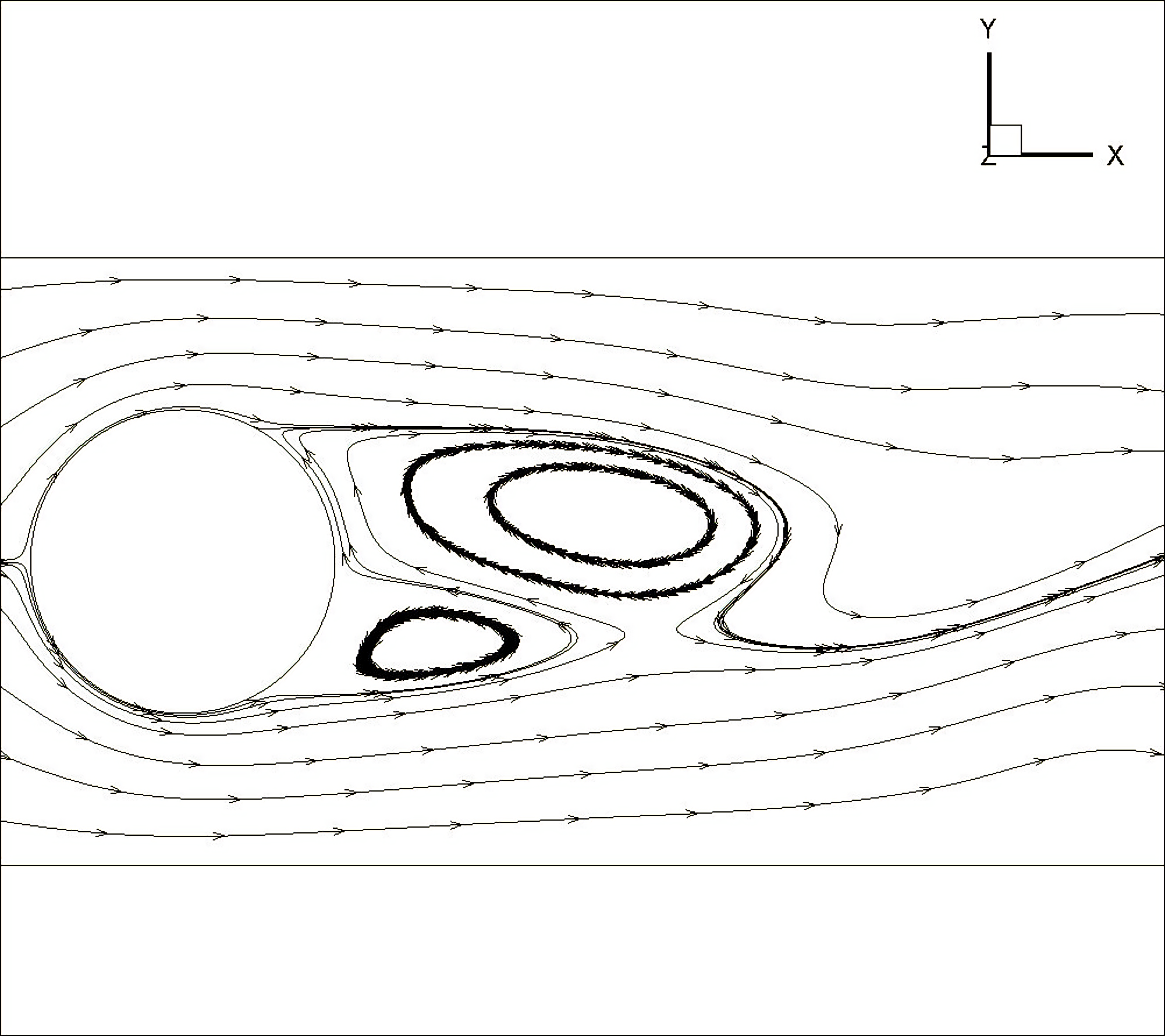}\label{fig:6b2}}\\
	\subfloat[$Re_{l}=219$, $n=1.4$]{\includegraphics[width=0.4\linewidth]{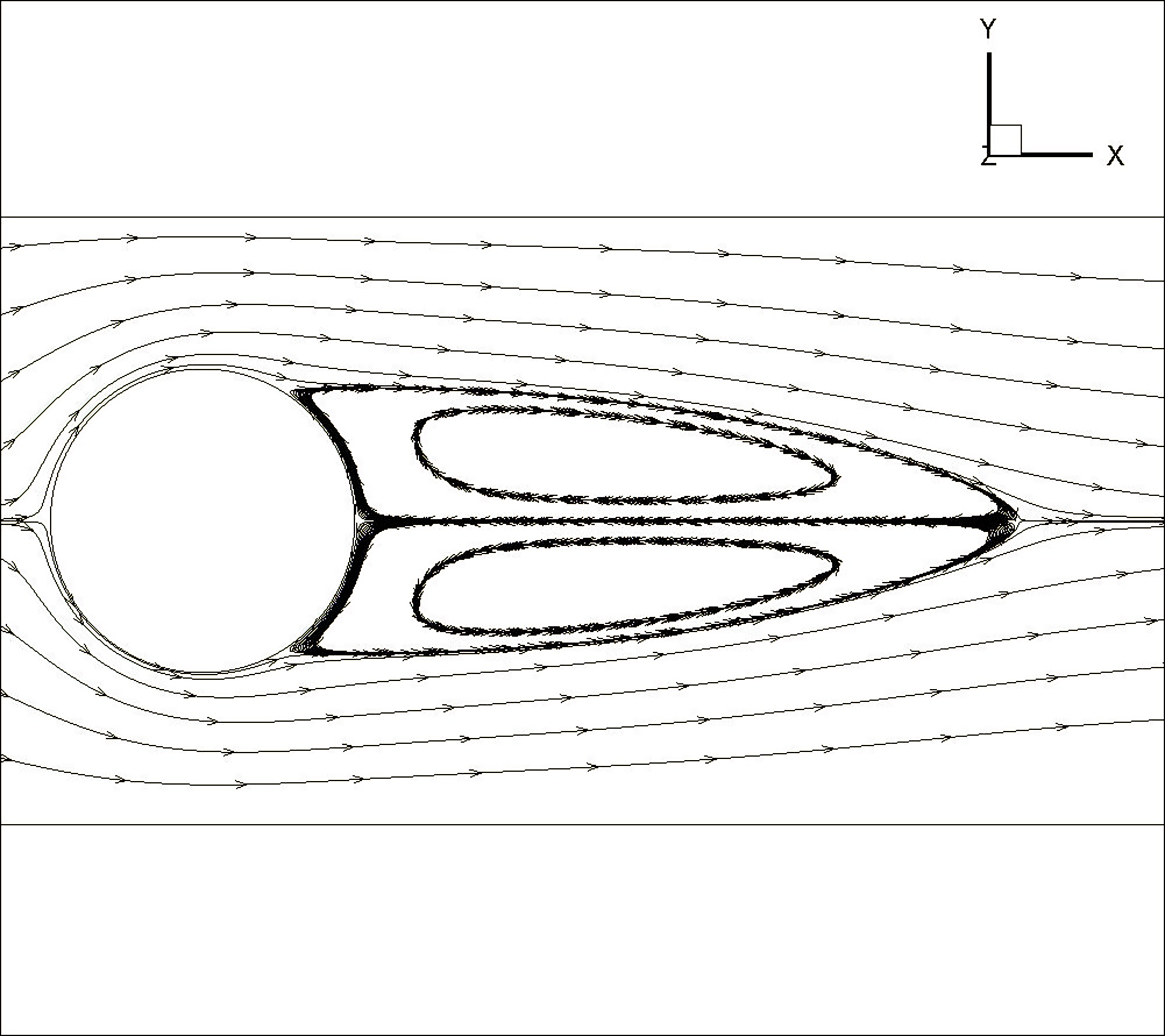}\label{fig:6c1}}\qquad
	\subfloat[$Re_{u}=220$, $n=1.4$]{\includegraphics[width=0.4\linewidth]{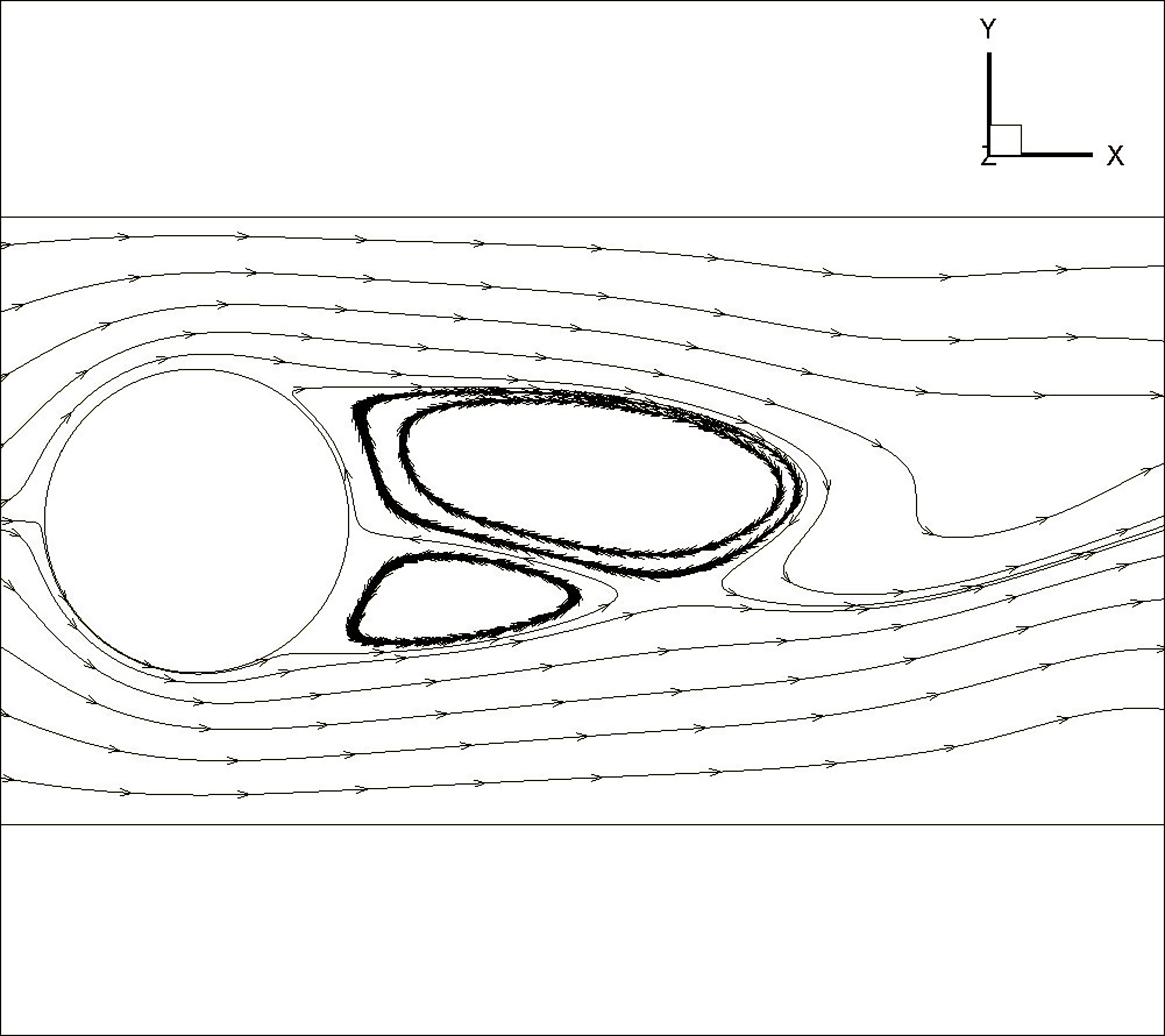}\label{fig:6c2}}\\
	%
	%
	%
	\subfloat[$Re_{l}=449$, $n=1.8$]{\includegraphics[width=0.4\linewidth]{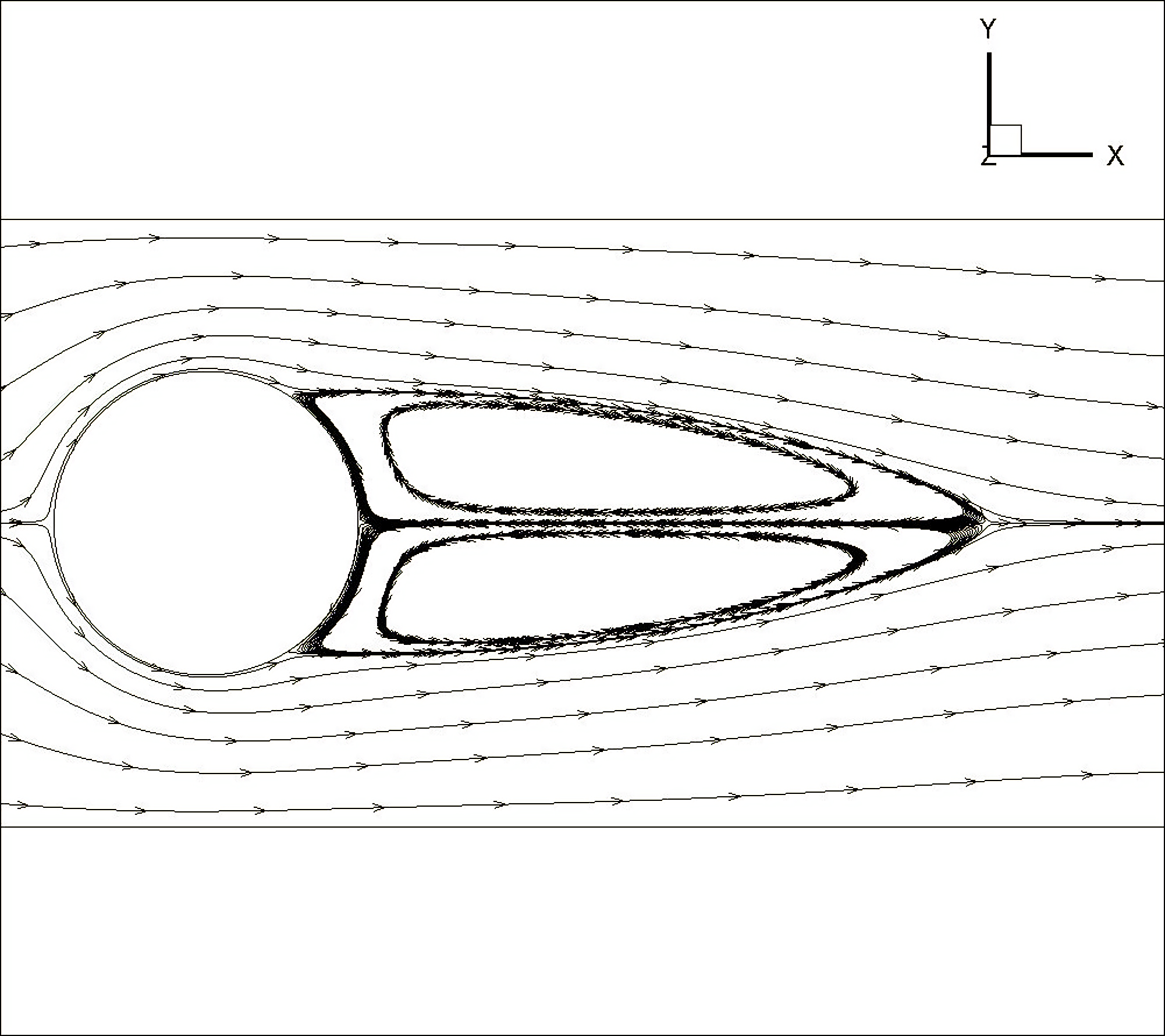}\label{fig:6e1}}\qquad
	\subfloat[$Re_{u}=450$, $n=1.8$]{\includegraphics[width=0.4\linewidth]{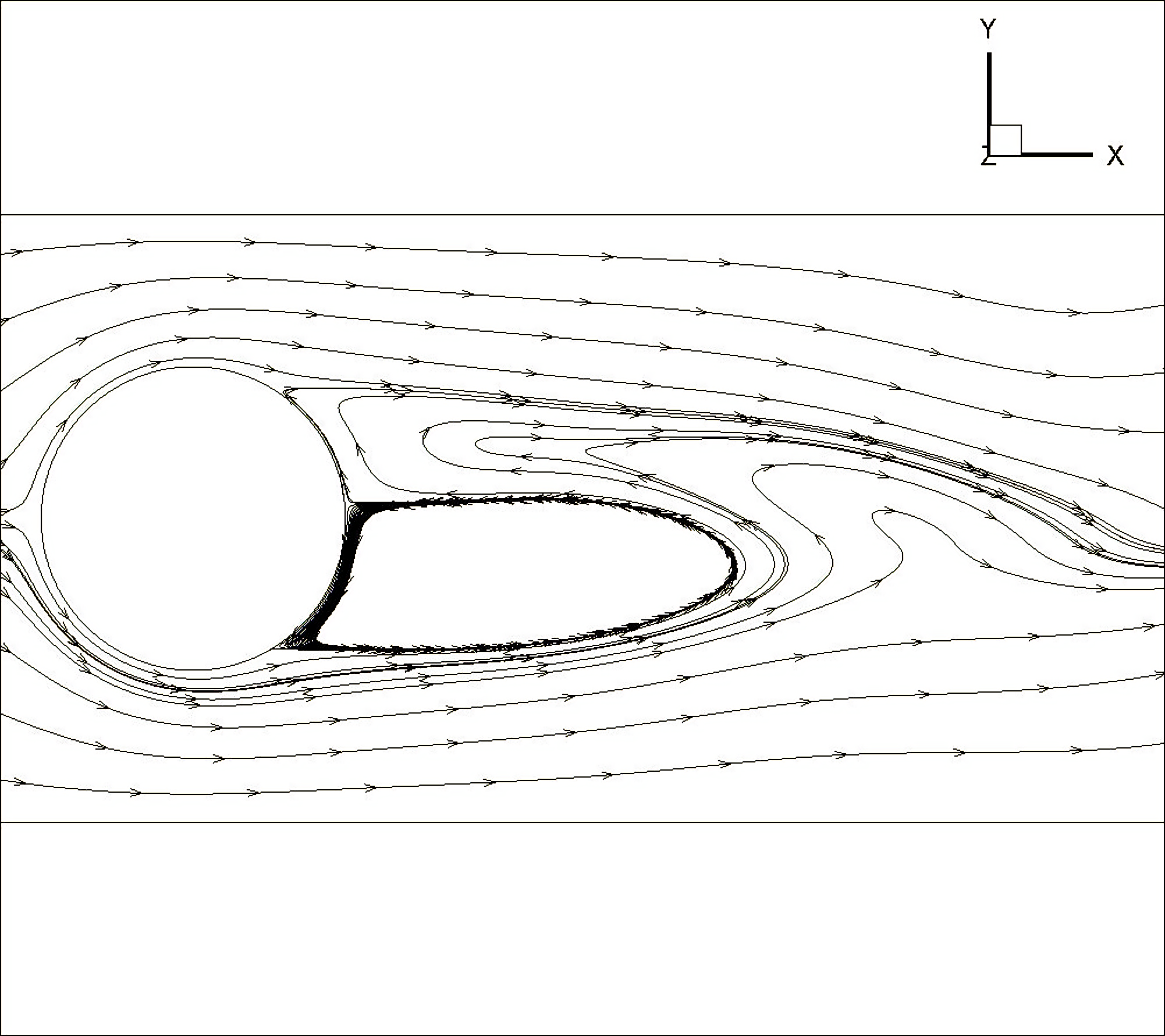}\label{fig:6e2}}\\
	\centering (g - l) $\beta=2$
\end{minipage}
%
\caption{Streamline profiles representing the upper critical Reynolds numbers ($Re_{l}\le Re_c\le Re_{u}$)  for various values of power-law index ($n$) for wall blockage $\beta=4$  and 2.
	The Reynolds numbers $Re_{l}$ and $Re_{u}$ indicate `symmetric wake flow' and `asymmetric wake flow', respectively.}
\label{fig:4}
\end{figure} 
\fig\ref{fig:4} displays the streamline profiles for the shear-thickening ($1\le n\le 1.8$) fluid flow over a channel confined cylinder for blockage ratio of $\beta=4$ and 2 at the two Reynolds number $Re_{\text{l}}$ and   $Re_{\text{u}}$ indicating the `symmetric' and `asymmetric' wake flow, respectively. The upper critical Reynolds number ($Re_{\text{l}}\le Re_{\text{c}}\le Re_{\text{u}}$) is thus marked as the lowest point or appearance for the two-dimensional asymmetric wake flow regime.  The dependence of the critical Reynolds number ($Re_{\text{c}}$) indicating the onset of wake instability on the power-law index ($n$) and the wall blockage ($\beta$) is presented in \tab\ref{Tab:4}. The results for unconfined  ($\beta=\infty$) flow are also \rev{obtained and} listed in  \tab\ref{Tab:4} for the comparison purpose\rev{, which are consistent with those reported elsewhere \citep{Sivakumar2006}}. 
Similar to the onset of flow separation (i.e., $Re^{\text{c}}$), the flow transition from symmetric to asymmetric wake formation also delays with strengthening of shear-thickening (i.e., increasing $n\ge 1$) behaviour of fluid for a given wall blockage ($\beta$). For a given fluid (i.e., fixed $n$), the transition of wake instability also delays with decreasing value of $\beta$ (i.e., increasing wall confinement).
For instance,  the critical Reynolds number ($Re_{\text{c}}$) value increases from $\sim 70.25$ to $\sim 179.5$ and from $\sim 84.5$ to $\sim 449.5$ with increasing value of power-law index ($n$) from 1 to 1.8 at blockage ratio ($\beta$) of 4 and 2, respectively. 
In contrast, $Re_{\text{c}}$ decreases with increasing value of power-law index ($n$) in unconfined ($\beta=\infty$) flow. For instance, $Re_{\text{c}}$ value decreases from $\sim 46.5$ to $\sim 33.5$ with increasing value of $n$ from 1 to 1.8, respectively.  These trends of the onset of wake instability are qualitatively consistent with the literature \citep{Sahin2004,Sivakumar2006,Bharti2007b} for the limiting conditions.
\\\noindent 
To broaden the usefulness of the results, the functional dependence of the critical Reynolds number ($Re_{\text{c}}$) on the power-law index ($n$) and blockage ratio ($\beta$) is expressed by the following predictive correlations (\eqn\ref{rec_nb}). 
\begin{equation}
	Re_c(n, \beta)= a_4n^4 + a_3n^3+a_2n^2+a_1n+ (a_0 \pm \Delta)	
	\qquad \text{for}\quad 2\le \beta \le \infty\quad\text{and}\quad 1\le n\le 1.8	
%
\label{rec_nb}
\end{equation}
\rev{Based on the statistical analysis of numerical data, the coefficients ($a_0$ to $a_4$ and $\Delta$) appearing in the above predictive correlation (\eqn\ref{rec_nb}) are noted in \tab\ref{Tab:5}.} 
The critical Reynolds number ($Re_{\text{c}}$) values, in general, have shown quartic (i.e., 4th order) dependence on the power-law index ($n$), irrespective of the wall interference ($\beta$).  While the functional dependence is even (i.e., $4^{\text{th}}$) degree polynomial of $n$ for all $\beta$, notably, the leading coefficients (i.e., $a_4/a_0$) is positive for confined ($\beta=2$ and 4) whereas negative for unconfined ($\beta=\infty$) flow.
The predictive expressions thereby suggest a decreasing value of $Re_{\text{c}}$ with increasing $n$ for $\beta=\infty$ and vice versa for $0 < \beta < \infty$.  
\\\noindent 
To further understand the effects of $n$ and $\beta$ \rev{on the onset of wake instability, $Re_{\text{c}}$ has been  normalized with respect to (i) an unconfined flow of non-Newtonian fluid ($\text{X}_{\text{c}}$), and (ii) an unconfined flow of Newtonian fluids ($\text{Y}_{\text{c}}$), as defined by  \eqn(\ref{nrec1})}.
\begin{equation}
\text{X}_{\text{c}}=\frac{Re_{\text{c}}(n,\beta)}{Re_{\text{c}}(n,\infty)}\qquad\text{and}\qquad
\text{Y}_{\text{c}}=\frac{Re_{\text{c}}(n,\beta)}{Re_{\text{c}}(1,\infty)}
\label{nrec1} 
\end{equation}
%
\begin{figure}[h]
	\centering
	%
	%
	\subfloat[$\text{X}_{\text{c}}(n,\beta)$]{\includegraphics[width=0.4\linewidth]{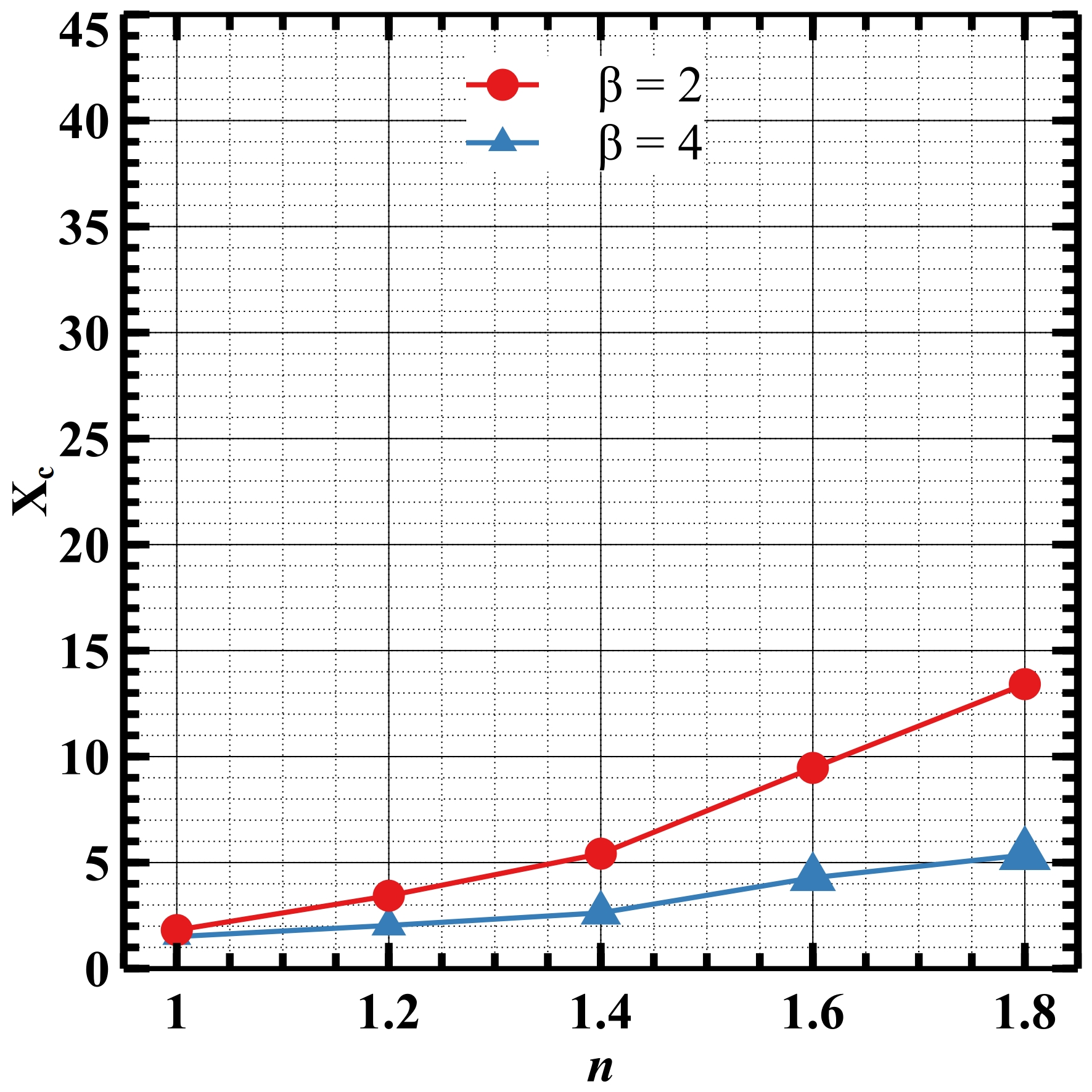}\label{fig:3c}}\qquad
	\subfloat[$\text{Y}_{\text{c}}(n,\beta)$]{\includegraphics[width=0.4\linewidth]{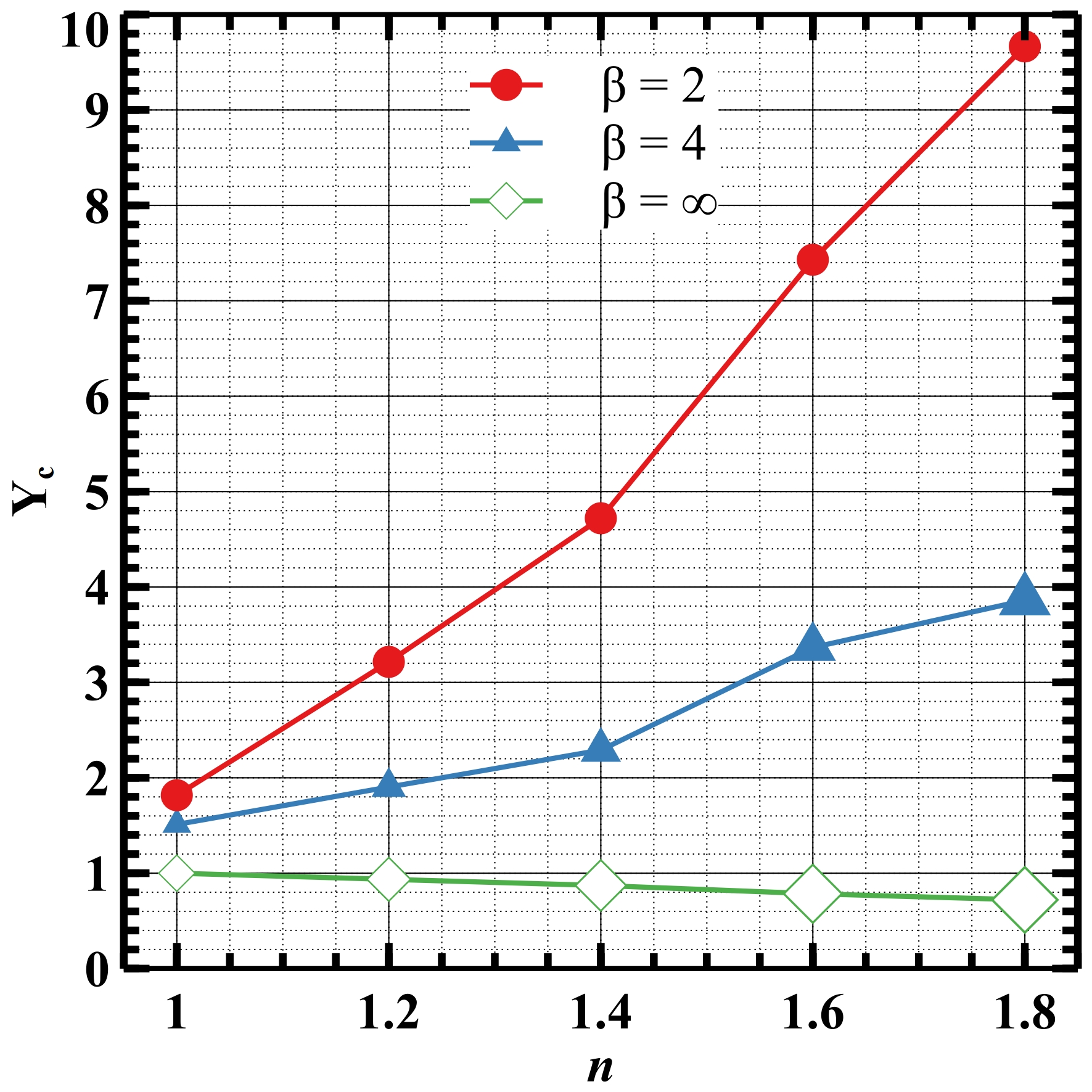}\label{fig:3d}}\\
	\caption{Normalized critical Reynolds numbers ($\text{X}_{\text{c}}$ and $\text{Y}_{\text{c}}$) as a function of power-law index ($n$) and wall blockage ($\beta$).}
	\label{fig:5}
\end{figure} 
%
\figs\ref{fig:3c} and \ref{fig:3d} display the complex dependence of the normalized factors ($\text{X}_{\text{c}}$ and $\text{Y}_{\text{c}}$) on the dimensionless parameters ($n$ and $\beta$).
The normalized ($\text{X}_{\text{c}}$ and $\text{Y}_{\text{c}}$) values have shown qualitatively similar dependence to that of critical Reynolds number ($Re_{\text{c}}$) on dimensionless parameters ($n$ and $\beta$).  The \rev{wake} transition, in comparison to unconfined flow, is delayed, i.e., $\text{X}_{\text{c}}$ increased with decreasing $\beta$ and increasing $n$. 
Similarly, in comparison to Newtonian fluid flow over unconfined cylinder, the transition is strongly delayed, i.e., $\text{Y}_{\text{c}}$ increased with decreasing $\beta$ and increasing $n$. 
The comparison of \figs \ref{fig:3} and \ref{fig:5}, however, shows that the blockage effects are stronger on onset of wake formation ($\text{X}^{\text{c}}$) than that on the onset of wake instability ($\text{X}_{\text{c}}$). The reverse trend is, however, observed with respect to the Newtonian flow over unconfined cylinder that  onset of wake instability ($\text{Y}_{\text{c}}$) is strongly influenced than that the onset of wake formation ($\text{Y}^{\text{c}}$). 
%
\rev{\begin{figure}[h]
	\centering
	\subfloat[$St_c$ vs $n$]{\includegraphics[width=0.45\linewidth]{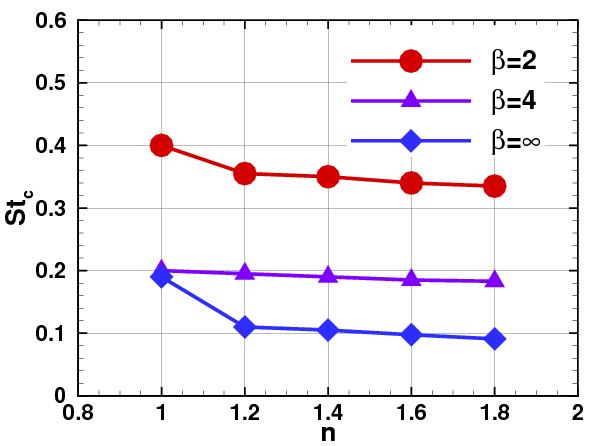}\label{fig:8a}}\qquad
	\subfloat[$C_{\text{D,c}}$ vs $n$]{\includegraphics[width=0.45\linewidth]{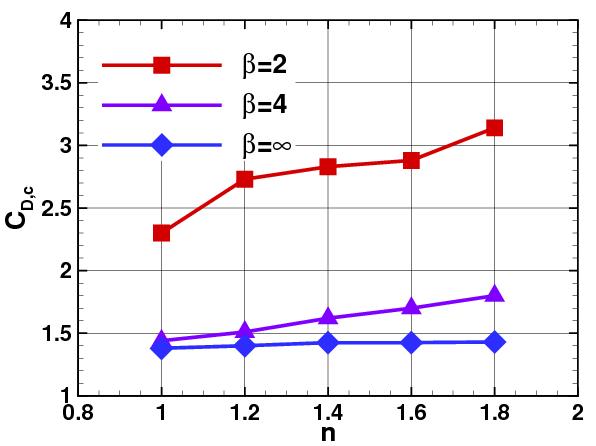}\label{fig:8b}}\\
	\caption{Dependence of critical (a) Strouhal number ($St_c$) and (b) time-averaged drag coefficient ($C_{\text{D,c}}$) on the power-law index ($n$) and wall blockage ($\beta$) at 
		critical Reynolds number ($Re_c$).}
	\label{fig:8}
\end{figure}} 
%
\rev{
\\\noindent
The frequency of vortex shedding is considered to be one of the prime characteristics of the asymmetric wake flows. It is presented herein in terms of the Strouhal number ($\mathit{St}_{\text{c}}$) at the critical condition. \fig\ref{fig:8} shows the Strouhal number ($\mathit{St}_{\text{c}}$) and total time-averaged drag coefficient ($C_{\text{D,c}}$) as a function of power-law index ($n$) and the blockage ratio ($\beta$) at the critical values of the Reynolds number ($Re_c$). 
Both $\mathit{St}_{\text{c}}$ and $C_{\text{D,c}}$ have increased with an increasing wall confinement ($\lambda$) whereas $\mathit{St}_{\text{c}}$ decreased and $C_{\text{D,c}}$ increased with increasing $n$, under otherwise identical conditions.
While the critical Reynolds number ($Re_c$) has shown complex dependence on $n$, i.e., increased for confined cylinder and decreased for unconfined cylinder, interestingly, the critical Strouhal number ($\mathit{St}_{\text{c}}$) decreased with increasing $n$, irrespective of the wall confinement ($\lambda = \beta^{-1}$). An increasing $\mathit{St}_{\text{c}}$ with increasing $\lambda$, irrespective of fluid behaviour ($n$), seems to be consistent with the existing literature \citep{Williamson1996,Zdravkovich1997Book,Zdravkovich2003Book} for Newtonian fluids, as $Re_c$ has increased with decreasing $\beta$.   
Such trends of $Re_c$ are attributed to the complex interplay of non-linear viscosity, flow suppression by the channel wall and increase in the flow velocity with decreasing flow area between the cylinder surface and channel wall with increasing channel confinement ($\lambda$).}
\\\noindent
In summary, both onsets of wake formation and  wake instability are influenced in complex manner with increasing degrees of shear-thickening ($n > 1$) and wall effects ($\beta < \infty$). The conceivable explanation for the preceding discussion can be given as follows. The flow field in the proximity of a cylinder depicts the complex interplay between the inertial, frictional, and pressure forces persisting in the fluid. These forces scale non-linearly and differently with characteristic velocity (${u}_{\text{avg}}$), power-law index ($n$), and characteristics length ($D$). In the present modeling framework, the frictional and inertial forces scale as $F_{\text{v}} \propto{u}_{\text{avg}}^n$ and $F_{\text{i}} \propto {u}_{\text{avg}}^2$, respectively. In case of shear-thickening ($n > 1$) fluids, the viscous force ($F_{\text{v}}$) grows whereas inertial force ($F_{\text{i}}$) remains unchanged with increasing value of power-law index ($n$) for the fixed velocity (${u}_{\text{avg}}$). In contrast, both forces $F_{\text{v}}$ and $F_{\text{i}}$ grow with increasing velocity (${u}_{\text{avg}}$) for a given fluid (i.e., fixed $n$), however, $F_{\text{i}} > F_{\text{v}}$ as $n < 2$. The relative influence of these two forces ($F_{\text{r}}=F_{\text{v}}/F_{\text{i}}\propto{u}_{\text{avg}}^{n-2}$) strengthen with decreasing fluid velocity (${u}_{\text{avg}}$)  and with increasing fluid behaviour index ($n$).  %
\rev{Additionally, the apparent viscosity ($\eta$) increases, above the Newtonian viscosity, with increasing shear rate ($\dot{\gamma}$) in shear-thickening ($n>1$) fluids. In turn, the viscous effects dominate over the inertial effects with increasing confinement ($\beta$ from $\infty$ to 2) even far away from the cylinder. In case of unconfined flow, the viscous effects dominate in the close vicinity of the cylinder whereas the inertial effects govern the flow far away from the cylinder \citep{Sivakumar2006}.}
Such a complex interactions of the forces yield non-monotonic trends seen in the preceding sections.   
The decrease in the critical Reynolds numbers ($Re^{\text{c}}$ and $Re_{\text{c}}$) with increasing power-law index ($n$) \rev{leading} to the `Stokes paradox' in unconfined flow is consistent with the above explanation. The far away boundaries have no impact on the wake formation/instability \rev{in unconfined flow}. 
An introduction of the wall blockage further accentuates these influences  due to coupled interaction of the additional hydrodynamic boundary layer developed on the walls and the fluid rheology. The hydrodynamic boundary layer thickness ($\delta_t$) is inversely proportional to the fluid velocity (${u}_{\text{avg}}$) as 
$\delta_t\propto {Re}^{-[1/(1+n)]}\approx {F_{\text{r}}}^{[1/(1+n)]}$ 
over the planar surface. In turn, the supplementary viscous force experienced by the fluid layers resist and stabilize the flow in complex manner due to non-linear variations of fluid velocity and viscosity in the boundary layer regions.  As the confinement increases (i.e., $\beta$ decreases), the flow remains stable for the larger range of velocity or Reynolds number. The confinement boundary therefore have stronger influence on the wake dynamics. The trends discussed in preceding sections have therefore shown an increase in the critical Reynolds numbers ($Re^{\text{c}}$ and $Re_{\text{c}}$) with increasing fluid behvaiour index ($n$) and decreasing blockage ratio ($\beta$). The stoke paradox observed in unconfined flows of power-law fluids would never be apparent in case of confined flows, under otherwise identical conditions.
%
\section*{Concluding remarks}
%
\noindent
In this work, the onsets of wake formation and instability are presented and discussed in terms of the critical Reynolds numbers ($Re^{\text{c}}$ and $Re_{\text{c}}$) for the two-dimensional flow of non-Newtonian \rev{shear-thickening} power-law fluid over a channel confined circular cylinder. The mathematical model equations have been solved using the finite volume method for a wide range of conditions, \rev{namely, power-law index} ($1\le n\le 1.8$), \rev{and wall blockage} ($\beta=2, 4$). Unconfined ($\beta=\infty$) flow results have also been obtained and presented for comparison purpose. The streamline ($\psi$), pressure ($C_{\text{p}}$), viscous ($C_{\text{f}}$), lift ($C_{\text{L}}$) and drag ($C_{\text{D}}$) coefficients profiles are analysed to obtain the critical conditions.  The following observations can be made from this work.
\begin{itemize}
	\item Both critical Reynolds numbers ($Re^{\text{c}}$ and $Re_{\text{c}}$) expressed complex dependence on the flow governing and influencing parameters ($n$ and $\beta$).
	\item Influence of power-law index ($n$) on {critical $Re$} is contrasting in confined ($\beta < \infty$) and unconfined ($\beta=\infty$) flows. \rev{With an increasing value of $n$, the critical Reynolds numbers ($Re^{\text{c}}$ and $Re_{\text{c}}$) increased in confined flow and decreased in unconfined flow.}
	\rev{The critical $Re$ values have increased with decreasing $\beta$ from $\infty$ to 2, irrespective of the fluid behaviour. For instance, $Re^{\text{c}}$ changed from 6.25 to 12.5 at $n=1$ and from 0.75 to 30.5 at $n=1.8$. Similarly, $Re_{\text{c}}$ altered from 46.5 to 84.5 at $n=1$ and from 33.5 to 449.5 at $n=1.8$.}  
	\item Both flow separation (i.e., wake formation) and asymmetry (i.e., wake instability) behind the cylinder delayed with increasing level of shear-thickening ($n$) and wall confinement ($\beta$).
	\item Wake length enhances but wake width suppresses with increasing $n$ and $\beta$. 
	\item Stokes paradox (i.e., no creeping flow), apparent in unconfined flow, is not relevant in confined flow \rev{of power-law fluids} over a cylinder.  
\end{itemize}
Finally, the predictive correlations for the critical $Re$ as a function of power-law index ($n$) and wall blockage ($\beta$) are presented for easy use in design and engineering of the relevant processes.
%
%
%
%
\section*{Declaration of Competing Interest}
\noindent 
All authors declare that they have no conflict of interest. The authors certify that they have NO affiliations with or involvement
in any organization or entity with any financial interest (such as honoraria; educational grants; participation
in speakers bureaus; membership, employment, consultancies, stock ownership, or other equity interest; and expert
testimony or patent-licensing arrangements), or non-financial interest (such as personal or professional relationships,
affiliations, knowledge or beliefs) in the subject matter or materials discussed in this manuscript.
\section*{Acknowledgments}
\noindent 
RPB duly acknowledge the 
Sponsored Research and Industrial Consultancy (SRIC), Indian Institute of Technology Roorkee, Roorkee (India) for providence of Faculty Initiation Grant (FIG), Ref. No. IITR/SRIC/886/F.I.G.(Scheme-A).  
%
%
\bibliographystyle{cas-model2-names}
%
\InputBibFiles{references}
%

\end{document}